\documentclass[aps,prl,twocolumn,groupedaddress,superscriptaddress,amsfonts,amssymb,amsmath,citeautoscript,a4paper]{revtex4-1}

\usepackage[utf8]{inputenc}
\usepackage[english]{babel}
\usepackage{microtype}
\usepackage{txfonts}
\usepackage{txfontsb}
\usepackage{bm}
\usepackage{xcolor}
\usepackage{graphicx}
\usepackage{float}
\usepackage{xspace}

\usepackage{enumerate}
\usepackage[inline]{enumitem}

\usepackage{siunitx}
\DeclareSIUnit[number-unit-product=]\percent{\char`\%} % remove space before percentage "units"

\usepackage[pdftex]{hyperref}
\hypersetup{colorlinks,
            linkcolor={blue!75!black!80!yellow},
            citecolor={blue!75!black!80!yellow},
            urlcolor={blue!75!black!80!yellow},
            pdfstartview=FitH}
            
\usepackage[eulergreek]{sansmath} % introduces the command \sansmath (analogous to \boldmath) to toggle sf-math in a paragraph (for caption)
\makeatletter
\renewcommand\@make@capt@title[2]{%
    \@ifx@empty\float@link{\@firstofone}{\expandafter\href\expandafter{\float@link}}%
    \sisetup{math-sf=\textsf}%
    \sansmath\sffamily\textbf{#1\@caption@fignum@sep}#2
}%

\makeatother

\newcommand{\ie}{i.e.\@\xspace}  %Gobble-spaces of the "small" type ("small" via \@)

\newcommand{\eg}{e.g.\@\xspace}

\usepackage{xifthen}
\usepackage{etoolbox}
\usepackage{soul}

\newcommand{\todo}[1]{% don't use in captions/subfloats etc.
	\textcolor{orange!80!yellow!95!black}{\textbf{[}%
	\ifthenelse{\isempty{#1}}%
		{\text{$\blacksquare$}}%
		{{\small\textsf{#1}}}%
	\textbf{]}}}

\usepackage{mdframed}
\definecolor{blue-violet}{rgb}{0.54, 0.17, 0.89}

\newmdenv[topline=false, rightline=false, bottomline=false,%
  linewidth=1.25pt, innerrightmargin=0pt, leftmargin=-8pt,%
  innerleftmargin=7pt, skipabove=0pt, skipbelow=0pt,%
  linecolor=blue-violet, fontcolor=blue-violet]{mdleftbar}

\newcommand{\ep}{\varepsilon}
\newcommand{\epd}{\ep_{\text{d}}}
\newcommand{\epm}{\ep_{\text{m}}}

\newcommand{\iu}{\text{i}}
\let\Re\relax\DeclareMathOperator{\Re}{Re}
\let\Im\relax\DeclareMathOperator{\Im}{Im}

\newcommand{\pll}{\parallel}
\DeclareMathOperator{\arctanh}{arctanh}

\newcommand{\wwp}{\omega_{\text{p}}}

\begin{document}

\title{Surface-response functions obtained from equilibrium electron-density profiles}

\author{N.~Asger~Mortensen}
\affiliation{\small Center for Nano Optics, University of Southern Denmark, Campusvej 55, DK-5230~Odense~M, Denmark}
\affiliation{\small Danish Institute for Advanced Study, University of Southern Denmark, Campusvej 55, DK-5230~Odense~M, Denmark}
\affiliation{\small Center for Nanostructured Graphene, Technical University of Denmark, DK-2800 Kongens Lyngby, Denmark}
\email{asger@mailaps.org}

\author{P.~A.~D.~Gon\c{c}alves}
\affiliation{\small Center for Nano Optics, University of Southern Denmark, Campusvej 55, DK-5230~Odense~M, Denmark}

\author{Fedor A. Shuklin}
\affiliation{\small Center for Nano Optics, University of Southern Denmark, Campusvej 55, DK-5230~Odense~M, Denmark}

\author{Joel~D.~Cox}
\affiliation{\small Center for Nano Optics, University of Southern Denmark, Campusvej 55, DK-5230~Odense~M, Denmark}
\affiliation{\small Danish Institute for Advanced Study, University of Southern Denmark, Campusvej 55, DK-5230~Odense~M, Denmark}

\author{Christos Tserkezis}
\affiliation{\small Center for Nano Optics, University of Southern Denmark, Campusvej 55, DK-5230~Odense~M, Denmark}

\author{Masakazu Ichikawa}
\affiliation{\small Department of Applied Physics, Graduate School of Engineering, The University of Tokyo, Bunkyo-ku, Tokyo 113-8656, Japan}

\author{Christian~Wolff}
\affiliation{\small Center for Nano Optics, University of Southern Denmark, Campusvej 55, DK-5230~Odense~M, Denmark}

\begin{abstract}
Surface-response functions are one of the most promising routes for bridging the gap between fully quantum-mechanical calculations and phenomenological models in quantum nanoplasmonics. Within all the currently available recipes for obtaining such response functions, \emph{ab initio} calculations remain one of the most predominant, wherein the surface-response function are retrieved via the metal's non-equilibrium response to an external perturbation. Here, we present a complementary approach where one of the most appealing surface-response functions, namely the Feibelman $d$-parameters, yield a finite contribution even in the case where they are calculated directly from the equilibrium properties described under the local-response approximation (LRA), but with a spatially varying equilibrium electron density. Using model calculations that mimic both spill-in and spill-out of the equilibrium electron density, we show that the obtained $d$-parameters are in qualitative agreement with more elaborate, but also more computationally demanding, \emph{ab initio} methods. The analytical work presented here illustrates how microscopic surface-response functions can emerge out of entirely local electrodynamic considerations.
\end{abstract}

\maketitle

\section{Introduction}

The plasmonic response of metallic nanostructures is commonly explored within the framework of classical electrodynamics~\cite{Jackson:1998}, while describing the free electrons of metals classically within the Drude-like local-response approximation (LRA)~\cite{Maradudin:2014}. This implies treating the electrons as a gas of noninteracting electrons, homogeneously distributed inside the metal and confined by a hard-wall at the metal's surfaces. In this fashion, any aspect of nonlocal (\ie, $q$-dependent) response~\cite{Barton:1979,Pitarke:2007,Raza:2015a} are commonly neglected both in the bulk of the metal (\eg, finite compressibility of the Fermi gas) and at its surface (\eg, Friedel oscillations and electronic spill-out associated with a finite work function).

\begin{figure}[t!]
    \centering
    \includegraphics[width=0.9\columnwidth]{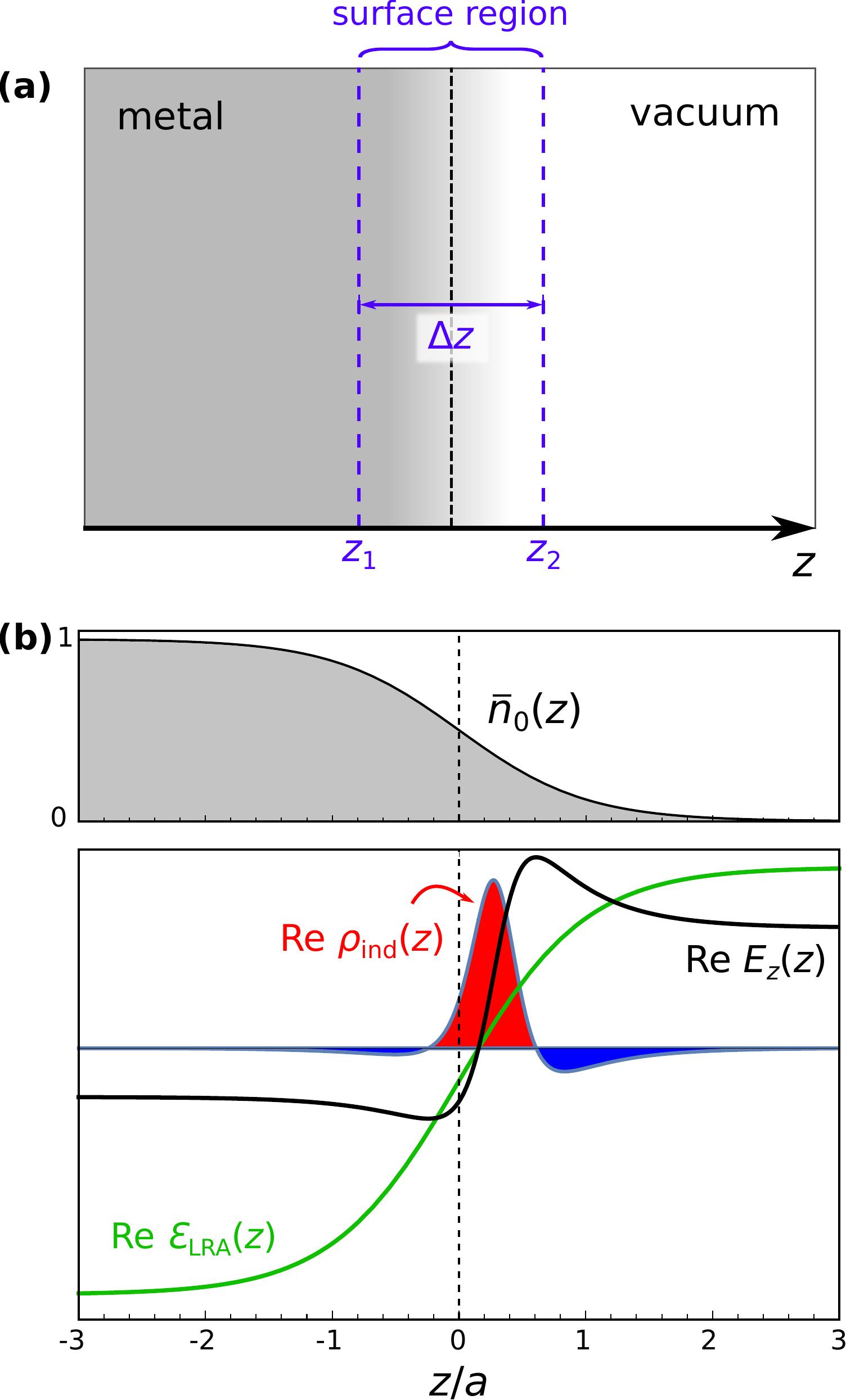}
    \caption{\textbf{a} Metal--vacuum interface, indicating the surface region where the electron density varies from its asymptotic, bulk values $\epm \equiv \varepsilon_\text{\textsc{lra}}(-\infty)$ and $\epd \equiv \varepsilon_\text{\textsc{lra}}(\infty) = 1$.
    \textbf{b} Top:  Schematic of the (normalized) equilibrium electron-density profile $\bar{n}_0(z)$ characterized by a smearing length $a$ in the vicinity of the surface (here defined by the $z=0$ plane). 
    Bottom: Real part of the system's dielectric function $\Re \varepsilon_\text{\textsc{lra}}(z)$ [Eq.~(\ref{eq:eps_Drude_z})] associated with $\bar{n}_0(z) = [1 - \tanh(z/a)]/2$, along with the ensuing $\Re E_z(z)$ and $\Re \rho_{\text{ind}}(z)$. All quantities are in arbitrary units. Parameters: $\omega=\wwp/\sqrt{3}$, and a Drude-type bulk damping of $\gamma/\wwp=0.3$
    \label{fig:fig1}}
\end{figure}

Despite its neglect of quantum-mechanical effects, the LRA has constituted a critical theoretical framework in the overall developments of plasmonics~\cite{Maradudin:2014,Stockman:2011,Fernandez-Dominguez:2017}. More recently, the importance of quantum phenomena has been pursued via both classical accounts, including smooth equilibrium electron-density profiles~\cite{Keller:1993,Ozturk:2011,David:2014}, and semiclassical hydrodynamic models~\cite{GarciadeAbajo:2008,Mortensen:2014,Toscano:2015}, as well as through \emph{ab initio} studies~\cite{Varas:2016,Zhu:2016}. The former approaches can be criticized for only dealing with some quantum aspects semiclassically, while the latter are typically by their complexity and by its practical applicability to small plasmonic systems~\cite{Zuloaga:2009,Zuloaga:2010,Teperik:2013,Andersen:2013,Sinha-Roy:2017}. 
In this context, surface-response functions aim to capture the dominant quantum phenomena and microscopic aspects of the surface, while still allowing for a (semi)classical treatment of the light--matter interactions in the bulk of the metal. As such, there has recently been a renewed interest in electrodynamic surface-response functions~\cite{Feibelman:1982,Liebsch:1997,Deng:2020} in the context of plasmon-enhanced light--matter interactions~\cite{Bozhevolnyi:2017a,Fernandez-Dominguez:2018,Goncalves_SpringerTheses,Goncalves:2020} and quantum plasmonics~\cite{Tame:2013,Fitzgerald:2016,Zhou:2019}, including, in particular, their importance for understanding plasmon--emitter interactions in nanoscale environments~\cite{Goncalves_SpringerTheses,Goncalves:2020}, and plasmon-enhanced interactions with two-dimensional (2D) materials~\cite{Goncalves_SpringerTheses,Goncalves:2020a}, as well as for the understanding of detailed spectral properties of plasmon resonances themselves~\cite{Apell:1983,Teperik:2013,Yan:2015,Christensen:2017,Yang:2019,Echarri:2020}. 

Traditionally, surface-response functions have been obtained through first-principle calculations of the electrodynamics of metal surfaces subjected by time-varying electric fields~\cite{Langreth:1984}, e.g., by employing time-dependent density-functional theory (TDDFT)~\cite{Varas:2016}, while they can in some cases also be analytically evaluated from semiclassical models, such as the hydrodynamic model~\cite{Feibelman:1975,Christensen:2017,Svendsen:2020}. In all cases, the common strategy has been to first evaluate the non-equilibrium response to obtain the induced charge density, $\rho_\text{ind}(\omega;\mathbf{r}_{\mathbf{\hat{n}}})$, and, from it, extract the surface-response function, e.g., the Feibelman $d_\perp$-parameter (corresponding to the centroid of induced charge density~\cite{Feibelman:1982}). 
Here, we explicitly show that even when considering equilibrium properties alone and a local-response approach, there is a finite contribution to the metallic surface-response functions provided that the (equilibrium) electron density varies smoothly from its bulk value deep inside the metal to zero near the metal's surface~\cite{Keller:1986,Ichikawa:2011,Ichikawa:2018} (as opposed to terminate abruptly at it). Such an approach, despite its simplicity and inherent limitation, could nevertheless facilitate new physical insights into the electrodynamic fingerprints associated with quantum spill-out/spill-in, without resorting to computationally demanding \emph{ab initio} methods.

\section{Results}

We consider a metallic nanostructure where $n_0(\mathbf{r})$ is the equilibrium electron density (see~Fig.~\ref{fig:fig1}a), which is spatially inhomogeneous in the vicinity of the metal's surface, possibly including, \eg, quantum spill-out and/or Friedel oscillations~\cite{Friedel:1952} due to a finite work function~\cite{Lang:1970}. In the presence of time-harmonic electromagnetic fields, the electrodynamics of the system is governed by the integro-differential wave equation
\begin{equation}
\label{eq:waveequation-nonlocal}
 \bm{\nabla}\times \bm{\nabla }\times \mathbf{E}(\mathbf{r})=\frac{\omega^2}{c^2}  \int \text{d}\mathbf{r}'\, \varepsilon(\mathbf{r},\mathbf{r}')\mathbf{E}(\mathbf{r}'),
\end{equation}
where $\omega$ is the angular frequency, $c$ is the speed of light in vacuum, and $\varepsilon(\mathbf{r},\mathbf{r}')$ is the nonlocal linear-response function, i.e., the (nonlocal) dielectric function of the quantum electron gas (here assumed to be isotropic, for the sake of simplicity). The microscopic and analytical understanding of $\varepsilon(\mathbf{r},\mathbf{r}')$ is in general limited to bulk considerations within the random-phase approximation (RPA) or the hydrodynamic model (HDM)~\cite{Lindhard:1954,Mermin:1970,Keller:1986,Pitarke:2007,Raza:2015a,Goncalves_SpringerTheses}.

\textit{Local-response approximation (LRA)}. In order to proceed with the nonlocal, integro-differential wave equation~(\ref{eq:waveequation-nonlocal}), it is common to invoke further approximations---in the context of plasmonics, the prevailing one being the na{\"i}ve LRA, epitomized by
\begin{subequations}
\begin{align}
    \varepsilon(\mathbf{r},\mathbf{r}')&\approx  \varepsilon_\text{\textsc{lra}}(\mathbf{r})
    \delta(\mathbf{r}-\mathbf{r}').
\end{align}
Here, the inherent finite-range nonlocal response of the electron gas is neglected in favor of a zero-range, \emph{local} response (mathematically represented by the Dirac delta function in the previous expression). Physically, this is equivalent to neglecting spatial dispersion represented by a finite wave vector dependence of the dielectric function~\cite{Pitarke:2007,Raza:2015a,Goncalves_SpringerTheses}, and thus ignoring, for instance, the finite dynamic compressibility of the electron gas~\cite{Pitarke:2007,Raza:2015a}. In spite of this---and as we show in what follows---some quantum aspects associated with an inhomogeneous electron gas (Fig.~\ref{fig:fig1}a), like electronic spill-out, can still be incorporated to some extent in the LRA. In particular, the LRA enables the simplification of the nonlocal wave equation~(\ref{eq:waveequation-nonlocal}) to the local-response one:
\begin{equation}
\label{eq:waveequation-LRA}
 \mathbf{\nabla}\times \mathbf{\nabla }\times \mathbf{E}(\mathbf{r})=\frac{\omega^2}{c^2} \varepsilon_\text{\textsc{lra}}(\mathbf{r})\mathbf{E}(\mathbf{r}),
\end{equation}
\end{subequations}
which is conceptually simpler and computationally more tractable~\cite{Gallinet:2015}.

\textit{Piecewise-constant approximation (PCA)}. Inspired by long-established traditions in the electrodynamics of composite dielectric problems~\cite{Joannopoulos:2008}, it is common in plasmonics~\cite{Maradudin:2014} to invoke yet another approximation: the step-like, abrupt surface termination of the metal, thereby neglecting any microscopic inhomogeneities in the vicinity of the surface (herein defined by $z=0$, without loss of generality, with the metal and the dielectric each occupying the $z < 0$ and $z > 0$ half-spaces, respectively). Under this approximation, $\varepsilon_\text{\textsc{lra}}(z) \to \ep_{\text{\textsc{pca}}}(z)$, with
\begin{align}
\label{eq:PCA}
    \ep_{\text{\textsc{pca}}}(z) &\equiv \varepsilon_\text{\textsc{lra}}(-\infty) \Theta(-z) + \varepsilon_\text{\textsc{lra}}(\infty) \Theta(z) \\[0.5em]
    &\equiv \ep_{\text{m}} \Theta(-z) + \ep_{\text{d}} \Theta(z) ,
\end{align}
where the system's dielectric function is constructed out of two interfacing piecewise-constant (bulk) local-response functions, $\epm \equiv \epm(\omega) $ and $\epd \equiv \epd(\omega)$ (and Eq.~(\ref{eq:waveequation-LRA}) is then solved by invoking the classical pillbox arguments at these interface~\cite{Jackson:1998}). Here, $\epm$ is the Drude-like dielectric function of the free-electron gas~\cite{Maradudin:2014,GarciadeAbajo:2008}
\begin{equation}
 \ep_{\text{m}} = \ep_{+} - \frac{\omega_{\text{p}}^2}{\omega^2 + \iu \omega \gamma} , 
 \label{eq:eps_Drude}
\end{equation}
with $\ep_{+} \equiv \ep_{+}(\omega)$ allowing for the incorporation of the polarization due to the positive ionic background or for a heuristic account of interband transitions. It should be emphasized that the PCA has been tremendously successful in advancing the field of plasmonics, being sufficient to interpret the majority of experimentally observed phenomena~\cite{Maradudin:2014}. What makes the PCA legitimate in most cases is the fact that the electron density is only non-uniform across an extremely small region in the vicinity of the metal surface, typically spanning only a few \r{a}ngstr\"{o}ms [\ie, on the order to the metal's Fermi wavelength (in the bulk), $\lambda_{\text{F}}$]. In spite of this, such a ``classical'', piecewise-constant approximation, is currently being challenged by the recent developments in nanoscale plasmonics and plasmon-empowered light--matter interactions at nanometric scales~\cite{Fernandez-Dominguez:2018,Zhu:2016,Dombi:2020,Yang:2019,Goncalves:2020,Goncalves:2020a,Goncalves_SpringerTheses}.

\textit{Surface-response functions.} In the PCA, the induced charge is strictly a (singular) surface charge, \ie, $\rho_\text{ind}(z) \propto \delta(z)$~\cite{Jackson:1998,Feibelman:1982,Goncalves_SpringerTheses}, while in reality, however, it assumes a nonsingular induced charge density $\rho_\text{ind}(z)$ of a finite, surface-peaked nature (Fig.~\ref{fig:fig1}b). In this context, the Feibelman $d$-parameters, $d_\perp \equiv d_\perp(\omega)$ and $d_\pll \equiv d_\pll(\omega)$, are dynamical surface-response functions that correspond to the first-moment (\ie, the centroid) of the induced charge density and of the normal derivative of the tangential current density, given, respectively, by ($\omega$-dependence implicit)~\cite{Feibelman:1982}
\begin{equation}
d_\perp = \frac{\int_{-\infty}^\infty \text{d}z\, z \, \rho_\text{ind}(z)}{\int_{-\infty}^\infty \text{d}z\, \rho_\text{ind}(z)} ,
\qquad 
d_\parallel = \frac{\int_{-\infty}^{\infty} \text{d}z\, z\, \frac{\partial}{\partial z} J_x^{\text{ind}} (z)}{\int_{-\infty}^{\infty} \text{d}z\, \frac{\partial}{\partial z} J_x^{\text{ind}}(z)} 
, \label{eq:Feibelman}
\end{equation}
which are complex-valued surface-response function, i.e., $d_\alpha(\omega)=d_\alpha'(\omega)+\iu d_\alpha''(\omega)$ with $\alpha \in \{\perp,\pll\}$. The general appeal of the $d$-parameters is that, once they are obtained, the system's optical response can be calculated by solving a $d$-parameter-modified classical electrodynamic problem, namely, the LRA wave equation~(\ref{eq:waveequation-LRA}) together with the ``classical'' PCA [recall Eq.~(\ref{eq:PCA})] but now subjected to the $d$-parameter-corrected, mesoscopic boundary conditions~\cite{Yan:2015,Christensen:2017,Yang:2019,Goncalves:2020,Goncalves_SpringerTheses}. Computationally, this is clearly more attractive than having to solve the more complex integro-differential problem typified by Eq.~(\ref{eq:waveequation-nonlocal}), while at the same time such reformulation into a quantum-informed ``classical-equivalent'' electrodynamic problem also paves the way for further analytical work~\cite{Christensen:2017,Goncalves:2020,Goncalves_SpringerTheses}. Naturally, different mechanism can be incorporated (together or separately) via the $d$-parameters, \eg, nonlocality, quantum spill-out/spill-in, Landau damping, etc~\cite{Feibelman:1982,Liebsch:1987}. In the following, we limit our consideration to the LRA contribution to the $d$-parameters emerging solely from a spatially varying dielectric function, \ie, $\varepsilon_\text{\textsc{lra}}(z)$. 

Alternatively to Eqs.~(\ref{eq:Feibelman}), the $d$-parameters can also be written in terms of surface integrals associate with the difference between the actual, microscopic fields and the classical, ``Fresnel'' fields stemming from the PCA~\cite{Feibelman:1982,Apell:1981,Forstmann:1986,Langreth:1989,Liebsch:1997}, specifically (see Supplementary Material): 
\begin{subequations}
 \begin{align}
  d_\perp &= -\frac{ \ep_{\text{d}} }{ \ep_{\text{m}} - \ep_{\text{d}} } 
  \int_{-\infty}^{\infty} \mathrm{d}z\,  \frac{E_z(z) - E_z^{\text{\textsc{pca}}}(z)}{E^{\text{\textsc{pca}}}_z(0^-) }
, \\[1.0em]
  d_\pll &= \frac{1}{ \ep_{\text{m}} - \ep_{\text{d}} } 
  \int_{-\infty}^{\infty} \mathrm{d}z\, \frac{ D_x(z) - D^{\text{\textsc{pca}}}_x(z) }{\ep_0 E^{\text{\textsc{pca}}}_x(0^-)} , 
 \end{align}%
\label{eq:Feib_diffFields}%
\end{subequations}%
where $E_{x,z}^{\text{\textsc{pca}}},D_x^{\text{\textsc{pca}}}$ are fields obtained within the classical, piecewise-constant approach. In the long-wavelength regime and to leading-order in $q|z_2-z_1|$, the Feibelman $d$-parameters~(\ref{eq:Feib_diffFields}) associated with a local, but smoothly varying dielectric function $\varepsilon_\text{\textsc{lra}}(z)$ can be written as~\cite{Bagchi:1979,Feibelman:1981,Apell:1982,Apell:1983a,Forstmann:1986} (see Supplementary Material)
\begin{subequations}
\begin{align}
 d_\perp &= \frac{1}{\epsilon_{\text{m}}^{-1} - \epsilon_{\text{d}}^{-1}} \int_{-\infty}^{\infty} \mathrm{d}z \, \left[ \varepsilon_\text{\textsc{lra}}^{-1}(z) - \varepsilon_\text{\textsc{pca}}^{-1}(z) \right] , \\[1.0em] 
 d_\pll &= \frac{1}{\epsilon_{\text{m}} - \epsilon_{\text{d}}} \int_{-\infty}^{\infty} \mathrm{d}z \, \left[ \varepsilon_\text{\textsc{lra}}(z) - \varepsilon_\text{\textsc{pca}}(z) \right] .
\end{align} \label{eq:d-params_LRA-PCA} %
\end{subequations}%
Equations~(\ref{eq:d-params_LRA-PCA}) unambiguously illustrate how $\varepsilon_\text{\textsc{lra}}(x)\neq \varepsilon_\text{\textsc{pca}}(x)$ contributes to a finite $d_\perp$ and $d_\parallel$. 
Naturally, in general, there will also be further contributions to the $d$-parameters stemming from the nonlocal response of the electron gas [e.g. treated within the nonlocal random-phase approximation (RPA) or the hydrodynamic model (HDM)]; nevertheless, it is important to emphasize that there is a nonzero contribution to the surface-response already within the LRA once the PCA is relaxed. In the following, we shall illustrate this in more detail with an elementary model that elucidates the physics---within the constraints associated with the LRA---of both spill-out and spill-in of the metal's electron density. Despite its inherent simplicity, the strength of the simple model adopted below lies also in its ability to render analytical results in closed-form.

\textit{Metal surface with a smoothly varying electron density.} As mentioned previously, a more realistic representation of a metal surface is to abandon the assumption of an infinitely sharp dielectric--metal interface and instead allow the metal's electron density to vary smoothly from its value deep inside the metal, $n_0^{\text{bulk}} \equiv n_0(z \to -\infty)$, to zero well inside the vacuum (Fig.~\ref{fig:fig1}). This can be modeled through a simple generalization~\cite{Feibelman:1975,Feibelman:1982,Ahlqvist:1982,Keller:1986,Ichikawa:2011,Ichikawa:2018} of Eq.~(\ref{eq:eps_Drude}), that is
\begin{equation}
 \varepsilon_{\text{\textsc{lra}}}(z) = \ep_{\infty}(z) - \frac{\omega_{\text{p}}^2}{\omega^2 + \iu \omega \gamma} \, \bar{n}_0(z) , 
 \quad \text{where} \quad \bar{n}_0(z)  = \frac{n_0(z)}{n_0^{\text{bulk}}} . 
 \label{eq:eps_Drude_z}
\end{equation}
where $n_0(\mathbf{r}) \equiv n_0(z)$ is the spatial profile of the equilibrium electron density and $n_0^{\text{bulk}} \equiv n_0(z \to -\infty)$ refers to its value deep inside the metal. Here, $\ep_{\infty}(z)$ takes into account the variation from the background polarization, subjected to the requirement that deep inside the metal (dielectric) it converges to the polarization due to the jellium background of positive ions, $\ep_{\infty}(z \to -\infty) = \ep_{+}$ (to the dielectric's permittivity $\ep_{\infty}(z \to +\infty) = \ep_{\text{d}}$). 
As a complementary perspective, this can also be interpreted as the common local response of the Drude kind, but with a spatially varying plasma frequency, $\omega_{\text{p}}(z) \equiv \omega_{\text{p}} \sqrt{\bar{n}_0(z)} $. In passing, we note that Eq.~(\ref{eq:eps_Drude_z}) has been used widely over the years, including Refs.~\citenum{Bagchi:1978,Keller:1993,Ozturk:2011,Liu:2017,Skjolstrup:2018,Skjolstrup:2019,Taghizadeh:2019,Rivacoba:2019}. Finally, we note how the PCA mathematically emerges upon replacing $\bar{n}_0(z)$ by a Heaviside function, i.e., $\bar{n}_0(z) \to \Theta(-z)$, corresponding to the classical, step-like termination of the equilibrium electron density.

\begin{figure}[t!]
    \centering
    \includegraphics[width=1\columnwidth]{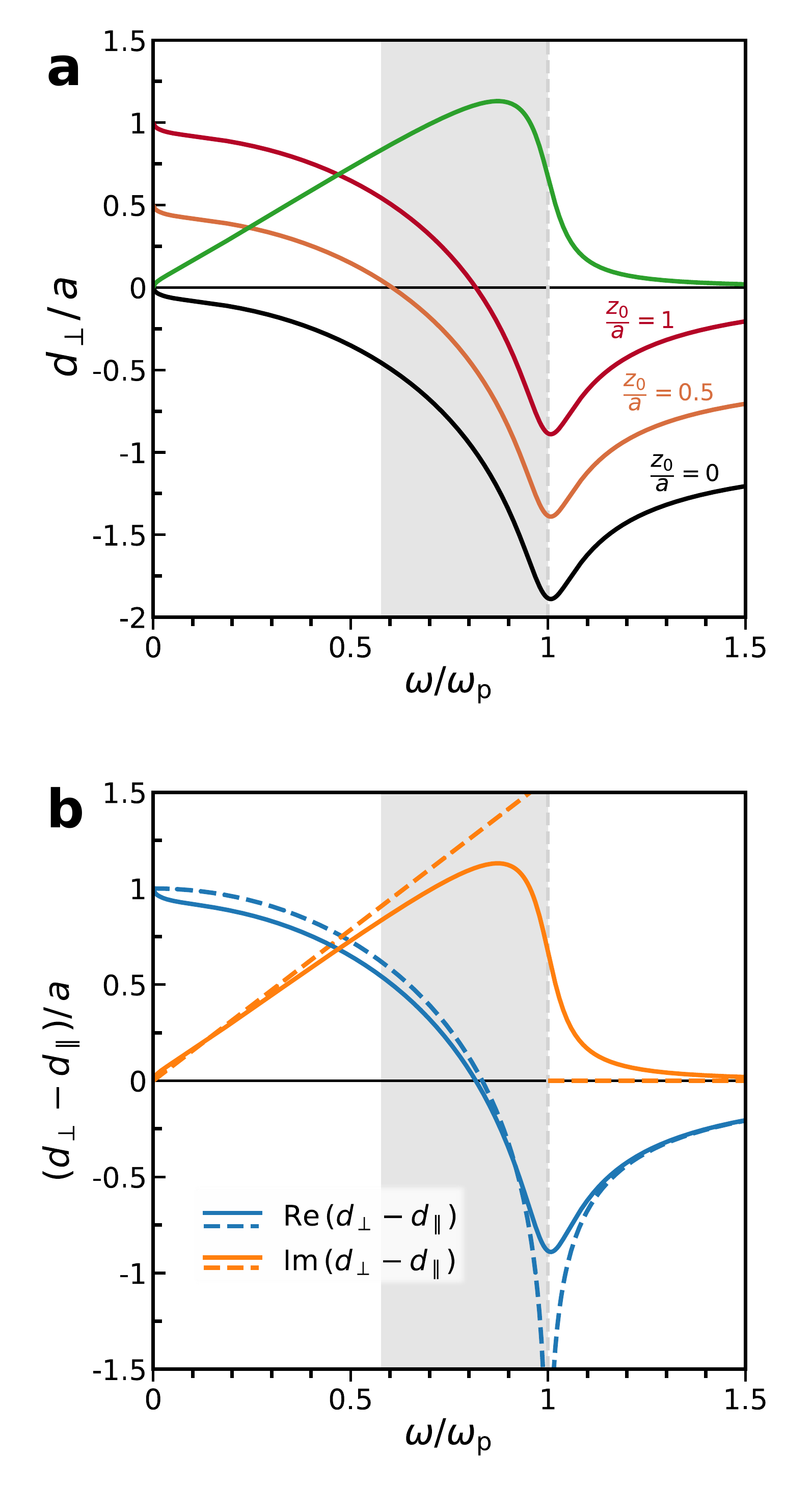}
    \caption{Feibelman $d$-parameters in the LRA for a jellium--vacuum interface ($\ep_+=\epd=1$) characterized by a smooth electron-density profile. 
    \textbf{a} Real, $\Re d_\perp$ (black, light-red, red), and imaginary part, $\Im d_\perp$ (green) [Eq.~(\ref{eq:spil-in-spill-out-a})], for the electron-density profile described in Eq.~(\ref{eq:tanh2}) with varying $z_0/a$; we assume a Drude bulk damping of $\Gamma=\gamma/\wwp=0.1$. 
    \textbf{b} Effective surface-response function $d_{\text{eff}} \equiv d_\perp-d_\pll$ [from Eq.~(\ref{eq:spil-in-spill-out})]. The dashed curves depict the result in the lossless case [Eq.~(\ref{eq:spil-in-spill-out-b}) and (\ref{eq:d_perp_simple_Gamma_0})]. The grey-shaded region indicates the frequency window supporting semiclassical localized plasmon resonances in metallic nanoparticles.}
    \label{fig:fig2}
\end{figure}

\textit{Transition from spill-in to spill-out.} To illustrate the transition from spill-in to spill-out, we consider a model electron-density profile of the form~\cite{Apell:1983a}
\begin{equation}
 \bar{n}_0(z) = \tanh^2 \left(\frac{z - z_0}{a} \right)\, \Theta(z_0 - z) 
 , \label{eq:tanh2}
\end{equation}
which is smooth and has the desired properties $\lim_{z \to -\infty} \bar{n}_0(z) = 1$ and $\lim_{z \to +\infty} \bar{n}_0(z) = 0$ [in fact, the latter can be made more stringent, \eg, $\lim_{z \to z_0} \bar{n}_0(z) = 0$]. The value $z_0$ indicates the position where the metal's electron density vanishes whereas the quantity $a$ characterizes the steepness of the spatial profile of the (normalized) equilibrium electron density [with $\lim_{a \to 0} \bar{n}_0(z) = \Theta(z_0-z)$]. 
The quantity $z_0$, in particular, governs whether the induced electron density spills inwards or outwards. For bulk electron-densities of typical plasmonic metals, both $a$ and $z_0$ amount to a few \r{a}ngstr\"{o}ms, and the model qualitatively captures the main results of self-consistent jellium considerations~\cite{Lang:1970}, while more refined models are needed to also represent finer details, \eg, Friedel oscillations~\cite{Friedel:1952,Rogowska:1994}.

Further, we assume that transition from the jellium background (i.e., the metal's positively charged ions) to the dielectric remains infinitely sharp because these only contain tightly bound electrons and thus are essentially immobile\footnote{We note, however, that this might not be the case for polar materials near its optical phonon frequencies.} when compared with the conductive (free-)electrons; hence, in the following we take 
\begin{equation}
 \ep_{\infty}(z) = \ep_{+} \Theta(-z) + \ep_{\text{d}} \Theta(z) , \label{eq:ep_infty_def}
\end{equation}
where we have assumed, without loss of generality, that the edge of jellium background is located at $z_{\text{b}}=0$.

\textit{Simple jellium next to vacuum.} For the of clarity, we first leave out background polarization effects or interband transitions and consider a simple jellium--vacuum interface, so that $\ep_+ = \epd = 1$. In this case, the integrals in Eqs.~(\ref{eq:d-params_LRA-PCA}) can be evaluated analytically, yielding
\begin{subequations}
\begin{align}
d_\perp(\Omega) &= z_0 - a\, \tilde{\Omega}\, \arctanh \left( \tilde{\Omega}^{-1}\right) 
, \label{eq:spil-in-spill-out-a} \\
d_\parallel(\Omega) &= z_0 - a. \label{eq:spil-in-spill-out-b}
\end{align}\label{eq:spil-in-spill-out}%
\end{subequations}%
where $\Omega = \omega/\omega_{\text{p}}$ and $\tilde{\Omega} = \sqrt{\Omega(\Omega + \iu \Gamma)}$, with $\Gamma = \gamma/\omega_{\text{p}}$. As we shall see, the frequency-independent result for $d_\parallel$ is a particular consequence of having assumed $\ep_+=\epd$. In the absence of bulk damping ($\Gamma \to 0^+$), Eq.~(\ref{eq:spil-in-spill-out-a}) can be written as~\cite{Apell:1983a}
\begin{equation}
 d_\perp(\Omega) = z_0 + a\, \frac{\Omega}{2} \left[ \ln \left| \frac{\Omega - 1}{\Omega + 1} \right|  + \iu \pi \Theta(1-\Omega)  \right] 
 , \label{eq:d_perp_simple_Gamma_0}
\end{equation}
with the low-frequency behavior of $d_\perp$ given by 
\begin{subequations}
\begin{align}
\Re d_\perp(\Omega \ll 1) &\simeq z_0 , \label{eq:spil-in-spill-out-asymptotic_Re} \\
\Im d_\perp(\Omega \ll 1) &\simeq a\, \frac{\pi}{2} \Omega . \label{eq:spil-in-spill-out-asymptotic_Im}
\end{align}\label{eq:spil-in-spill-out-asymptotic}%
\end{subequations}%
Notice that, even in the absence of \emph{bulk} damping, there is a nonzero contribution of \emph{surface-assisted} damping embodied through $\Im d_\perp \neq 0$ [see Eq.~(\ref{eq:d_perp_simple_Gamma_0})]. More fundamentally, this is a consequence of Kramers--Kronig relations (wherein a dispersive $\Re d_\perp$ renders $\Im d_\perp \neq 0$)~\cite{Dethe:2019}.
Moreover, we emphasize that the asymptotic limits~(\ref{eq:spil-in-spill-out-asymptotic}) are in agreement with results emerging from sum-rule considerations~\cite{Persson:1983,Persson:1985}. Interestingly, in the above result, $z_0$ coincides with the so-called static image-plane position that emerges from a self-consistent solution of the jellium perturbed by a static field~\cite{Lang:1973,Persson:1983,Persson:1985}, being a quantity of interest in surface science at large (a particular example being that of the surface-assisted van der Waals interaction of an atom near a metallic surface~\cite{Persson:1983,Apell:1981}). Recently, acoustic graphene plasmons have been proposed as a means to probe the quantum surface-response of metals~\cite{Goncalves:2020a} by placing a graphene sheet separated from a metal surface by a nanometric gap~\cite{Goncalves:2020b,Reserbat-Plantey:2021}. In particular, the static surface-response, $d_\perp(0)$ [which, within our simple treatment here, amounts to $z_0$; see Eq.~(\ref{eq:spil-in-spill-out-asymptotic_Re})], dependence could be experimentally probed in this way~\cite{Goncalves:2020a}.

The results [Eqs.~(\ref{eq:spil-in-spill-out})--(\ref{eq:d_perp_simple_Gamma_0})] for the a simple jellium surface next to vacuum are shown Fig.~\ref{fig:fig2}, showing how $\Re d_\perp$ is always negative for $z_0 = 0$ (Fig.~\ref{fig:fig2}a; black curve). Increasing $z_0/a$ brings the low-frequency part of $\Re d_\perp$ to positive values (Fig.~\ref{fig:fig2}a; light-red and red curves), potentially extending into the frequency regime $\wwp/\sqrt{3} \leq \omega < \wwp$ supporting semiclassical (specifically, within the HDM) localized surface plasmon (LSP) resonances in metal nanoparticles~\cite{Christensen:2014}. Consistent with causality and Kramers--Kronig relations, the dispersiveness of $\Re d_\perp$ is accompanied by a finite $\Im d_\perp$ (green, Fig.~\ref{fig:fig2}a; orange, Fig.~\ref{fig:fig2}b)~\cite{Dethe:2019,Persson:1983,Persson:1985}.

\begin{figure}[t!]
    \centering
    \includegraphics[width=\columnwidth]{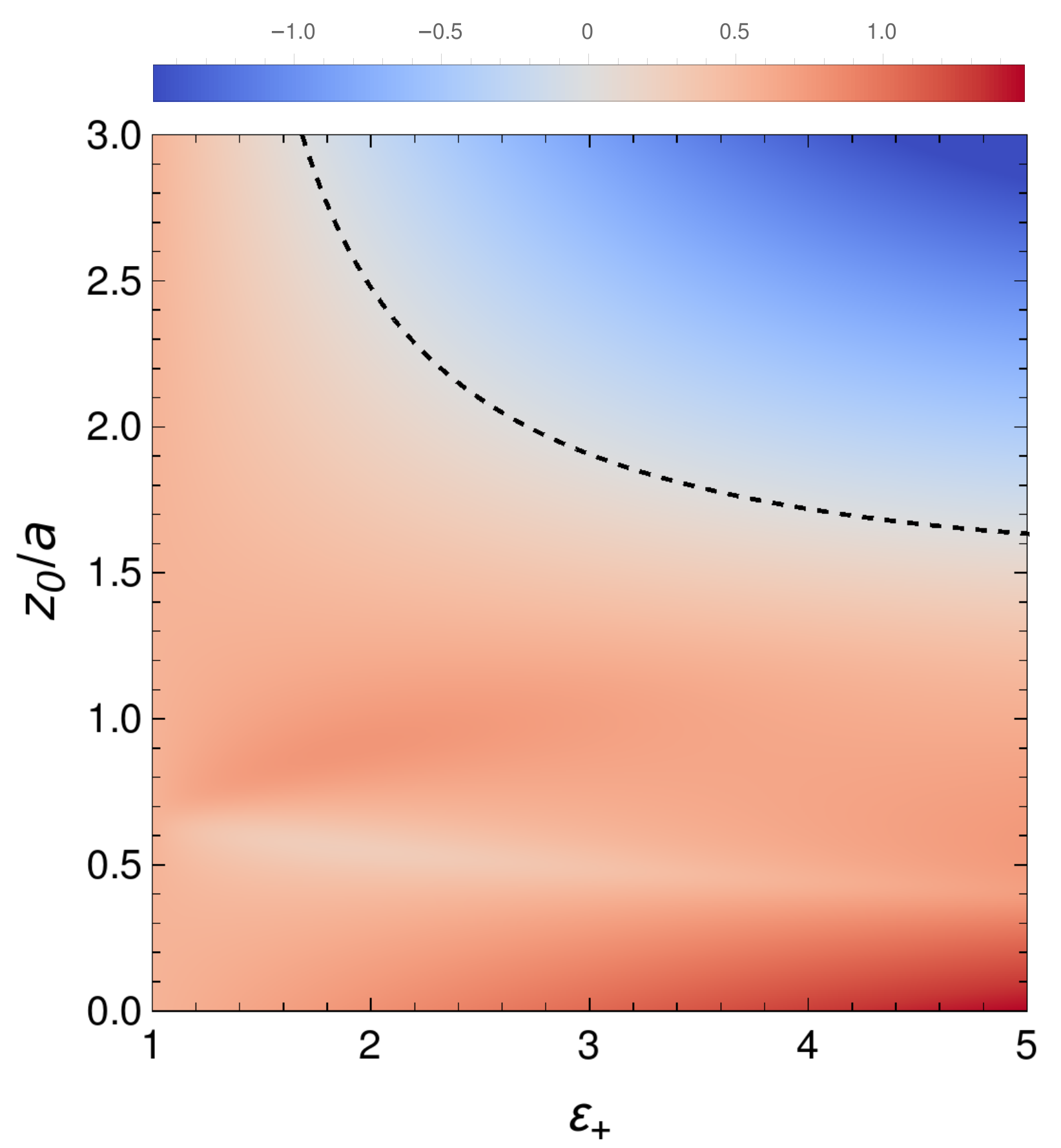}
    \caption{Density plot of $\Re \big(d_\perp-d_\parallel\big)/a$ in the ($\ep_+$,$ z_0$)-parameter space, computed at the classical quasistatic dipole LSP resonance frequency, $\omega=\wwp/\sqrt{\ep_+ + 2}$, of a spherical particle of radius $R$. The black dashed line indicates $\Re \big(d_\perp-d_\parallel\big) = 0$, thus separating regimes with nonclassical $1/R$ size-dependent spectral redshifts [reddish regions; $\Re \big(d_\perp-d_\parallel\big) > 0$] from blueshifts [bluish regions; $\Re \big(d_\perp-d_\parallel\big) < 0$]. We have assumed: $\epd=1$ and $\gamma/\wwp=0.1$.}
    \label{fig:fig3}
\end{figure}

\emph{Dipolar resonance of a metallic nanosphere.} To illustrate how the surface-response functions $d_\perp$ and $d_\pll$ jointly influence the optical response of a metallic nanostructure (Fig.~\ref{fig:fig2}b), we consider the prototypical case of a spherical nanoparticle of radius $R$; for simplicity, we take $\ep_+=1$ and assume that the nanosphere is in vacuum ($\epd=1$). Within the classical quasistatic LRA-description the spectrum of LSP resonances is dominated by a size-independent dipole resonance at the frequency $\omega=\wwp/\sqrt{3}$~\cite{Maradudin:2014,Wang:2006,Christensen:2014}. Accounting for nonclassical surface effects in a generalized Clausius--Mossotti relation,  the pole associated with the dipolar LSP resonance is, to leading-order in $d_{\perp,\pll}/R$, given by~\cite{Christensen:2017,Goncalves:2020,Goncalves_SpringerTheses} 
\begin{equation}
\label{eq:dipolecondiction}
0 = \epm + 2 
 - \left( \epm - 1 \right) \frac{2 \left(d_\perp-d_\parallel \right)}{R},
\end{equation}
which illustrates how the smearing of the jellium near the surface of the particle causes nonclassical $a/R$ size-dependent redshifts of the classical dipole resonance frequency (Fig.~\ref{fig:fig2}b). Crucially, in this case, \ie, with $\ep_+=\epd=1$, the ``effective'' surface-response function $d_{\text{eff}} \equiv d_\perp - d_\pll$~\cite{Liebsch:1997,Christensen:2017,Goncalves_SpringerTheses} has a ``universal'' behavior, namely, it is 
(i) independent of $z_0$, and 
(ii) proportional to the smearing of the spatially varying electron-density profile, characterized by the length $a$. Thus, interestingly, this indicates that, independently of $z_0$, the smearing itself contributes to a net nonclassical redshift ($\Re d_{\text{eff}} > 0$; spill-out) of the dipolar LSP resonance position of a jellium nanosphere in vacuum.

In the following, we simultaneously relax the assumptions of $\epd=1$ and of $\ep_+=1$. Allowing the latter to be larger than unity is commonly used to heuristically incorporate semiclassical accounts of background polarization effects or contributions arising from interband transitions in noble metals~\cite{Maradudin:2014,GarciadeAbajo:2008}.

\textit{Background and dielectric screening contributions.} Turning to the general case of  arbitrary $\ep_+$ and $\epd$, the effort required to perform the integrals~(\ref{eq:d-params_LRA-PCA}) are somewhat more elaborate, but can nevertheless still be evaluated analytically, reading (assuming $z_0 \geq 0$)
\begin{widetext}
\begin{subequations}
\begin{align}
 \frac{d_\perp}{a} &= C_\perp\, \left\{
 \frac{1 - \tilde{\Omega}^2 \ep_+}{1 - \tilde{\Omega}^2 \epd} \frac{z_0}{a} 
 - \frac{\epd}{\sqrt{\ep_+}} \tilde{\Omega} \left[  \arctanh\left(\frac{\tilde{\Omega}^{-1}}{\sqrt{\ep_+}}\right) 
 - \arctanh\left(\frac{\tilde{\Omega}^{-1}}{\sqrt{\ep_+}}\tanh\left(\frac{z_0}{a}\right) \right)
 \right]
  - \frac{\epm\sqrt{\epd}}{\epm + (\epd - \ep_+)} \tilde{\Omega}\,  \arctanh\left(\frac{\tilde{\Omega}^{-1}}{\sqrt{\epd}}\tanh\left(\frac{z_0}{a}\right) \right)
 \right\} 
 , \label{eq:spil-in-spill-out_general_perp} \\[1.0em] 
 \frac{d_\pll}{a} &= C_\pll\, \frac{z_0 - a}{a} 
 , \label{eq:spil-in-spill-out_general_pll}
\end{align}\label{eq:spil-in-spill-out_general}%
\end{subequations}%
\end{widetext}
where $C_\perp \equiv \big[ 1 + (\epd - \ep_+) \tilde{\Omega}^{2} \big]^{-1}$ and $C_\pll \equiv (\ep_{\text{m}} - \ep_{+})(\ep_{\text{m}} - \ep_{\text{d}})$, both being resonantly enhanced in the vicinity of $\omega=\wwp/\sqrt{\ep_+ - \epd}$; this Bennett-type resonance~\cite{Bennett:1970,Tsuei:1990} should not be confused with the common surface plasmon resonance occurring at $\omega=\wwp/\sqrt{\ep_+ + \epd}$. Moreover, contrasting with the previous case (where $\ep_+=\epd=1$), now \emph{both} $d_\perp$ and $d_\parallel$ are dispersive (\ie, exhibit frequency dependence). 

Finally, we note that these factors reduce to $C_\perp=C_\pll=1$ in the $\ep_+=\epd$ case. Additionally, in this particular case, $d_\perp$ and $d_\pll$ are given by Eqs.~(\ref{eq:spil-in-spill-out}) upon replacing $\tilde{\Omega}^{-1} \to \tilde{\Omega}^{-1}/\sqrt{\ep}$, where $\ep \equiv \ep_+ = \epd$. 

Returning to our discussion associated with Eq.~(\ref{eq:dipolecondiction}), we note that, in addition to the nonclassical $a/R$-dependent redshift of the resonance frequency, the $\propto (d_\perp-d_\parallel)/R$ term emerging in the pole of the polarizability~\cite{Goncalves:2020,Christensen:2017} [the generalized version of Eq.~(\ref{eq:dipolecondiction}) for arbitrary $\ep_+$ and $\epd$] now acquires a finite contribution also from $z_0$, which may lead to a net blueshift of the dipole LSP resonance. This is also in-line with recent experimental observations of the dependence of quantum size-effects on the local dielectric environment of the interface~\cite{Campos:2019}. As illustrated in Fig.~\ref{fig:fig3}, the combined effects of a non-unity interband permittivity, $\ep_+$, and of a finite $z_0$ may render the redshift of the classical dipole LSP resonance frequency into a net blueshift, depending on both $\ep_+$ and $z_0/a$ (and also on the particular value of the bulk-damping parameter, $\gamma$, which ``softens'' the sharp feature at $\omega=\omega_{\text{p}}$; see Fig.~\ref{fig:fig2}b). In this way, the model conceptually explains how different metals may exhibit contrasting $1/R$ size-dependencies of their surface plasmon resonances~\cite{Liebsch:1993,Liebsch:1987,Liebsch:1997}, towards the blue for $d_{\text{eff}} < 0$ (spill-in) and toward the red for $d_{\text{eff}} > 0$ (spill-out). An example of the former is silver (characterized by significant interband and valence band screening contributions to the optical response)~\cite{Liebsch:1997,Scholl:2012,Echarri:2020}, while an example of the latter is sodium (whose optical response is well described by a simple jellium treatment)~\cite{Liebsch:1997}. The imaginary part $\Im \big(d_\perp-d_\parallel\big)$ is a source of nonclassical $1/R$ size-dependent broadening~\cite{Christensen:2017,Goncalves:2020}. For the experimental visibility of nonclassical size-dependent shifts, it is naturally preferable that $\lvert \Re\big( d_\perp-d_\parallel\big)\rvert \gg \Im\big(d_\perp-d_\parallel\big)$, so that the nonclassical spectral shift is not rendered unobservable due to nonclassical damping.

\section{Discussion and Conclusions}

In this Article, we have revisited the concept of surface-response functions, highlighting that a finite contribution to the Feibelman $d$-parameters emerges even in a LRA-treatment with a spatially varying equilibrium electron-density profile---see Eq.~(\ref{eq:d-params_LRA-PCA}). While this insight has appeared in some form in the early literature~\cite{Bagchi:1979,Feibelman:1981,Apell:1982,Apell:1983a}, it has seemingly remained unnoticed in the more recent revival of surface-response functions and the widespread use of \emph{ab initio} accounts for quantum plasmonics. In working out this equilibrium contribution to the dynamic surface-response functions, we have deliberately omitted nonlocal corrections. In this context, the bulk nonlocal hydrodynamic response associated with the quantum compressibility of the electron gas (deliberately left out of our considerations) would contribute with a negative $\Re d_\perp$ (well below the plasma frequency, and for a jellium--vacuum interface), namely $ d_\perp = -\beta / \big( \wwp^2 - \omega^2 \big)^{1/2}$ and $d_\parallel=0$ ~\cite{Feibelman:1982,Goncalves:2020,Christensen:2017,Svendsen:2020}, with $\beta \propto v_{\text{F}}$~\cite{Halevi:1995,Raza:2015a,Goncalves_SpringerTheses} being a characteristic velocity of longitudinal plasmons. Qualitatively, this could enhance regimes in Fig.~\ref{fig:fig3} with a net blueshif, while consequently also reducing the spectral shift in regimes with a net redshift. This possible interplay of quantum compressibility and quantum spill-out is manifested in  self-consistent hydrodynamic treatments~\cite{Toscano:2015,Yan:2015b,Ciraci:2016}.

In conclusion, our analytical solution of the electrodynamics at metal surfaces transparently and unambiguously illustrates how the microscopic surface-response functions have a finite contribution originating entirely from equilibrium and local-response considerations. We believe that this is important insight for the understanding and further advancement of first-principle methods for the computation of accurate surface-response functions, as well as for the experimental exploration of mesoscopic optical phenomena at metal surfaces~\cite{Scholl:2012,Raza:2013,Raza:2015,Campos:2019,Yang:2019}. The latter is now becoming even more tangible with the advent of ultraconfined acoustic graphene plasmons
~\cite{Lundeberg:2017,Iranzo:2018,Dias:2018,Goncalves:2020a,Goncalves_SpringerTheses,Goncalves:2020b}.
Beyond the fundamental interest in surface-response functions, we note that the underlying quantum nonlocal response of the metals should also pose fundamental limitations for many light--matter interaction phenomena, ranging from surface-enhanced Raman spectroscopy~\cite{Toscano:2012b} to the perfect lens~\cite{Larkin:2005}.

{\small

\section{Acknowledgments.} 

This paper is dedicated to Mark~I.~Stockman in appreciation of his pioneering contributions to the broad area of Nano Optics. 
We thank T.~Christensen for valuable discussions and P.~M. Frederiksen for facilitating the writing of the manuscript.
N.~A.~M. is a VILLUM Investigator supported by VILLUM FONDEN (Grant No.~16498) and Independent Research Fund Denmark (Grant No.~7026-00117B).
J.~D.~C. is a Sapere Aude research leader supported by Independent Research Fund Denmark (Grant No.~0165-00051B).
C.~W. acknowledges funding from a MULTIPLY fellowship under the Marie Sk\l{}odowska-Curie COFUND Action (grant agreement No.~713694).
The Center for Nano Optics is financially supported by the University of Southern Denmark (SDU~2020 funding).
The Center for Nanostructured Graphene is sponsored by the Danish
National Research Foundation (Project No.~DNRF103).
}


\begin{thebibliography}{89}%
\makeatletter
\providecommand \@ifxundefined [1]{%
 \@ifx{#1\undefined}
}%
\providecommand \@ifnum [1]{%
 \ifnum #1\expandafter \@firstoftwo
 \else \expandafter \@secondoftwo
 \fi
}%
\providecommand \@ifx [1]{%
 \ifx #1\expandafter \@firstoftwo
 \else \expandafter \@secondoftwo
 \fi
}%
\providecommand \natexlab [1]{#1}%
\providecommand \enquote  [1]{``#1''}%
\providecommand \bibnamefont  [1]{#1}%
\providecommand \bibfnamefont [1]{#1}%
\providecommand \citenamefont [1]{#1}%
\providecommand \href@noop [0]{\@secondoftwo}%
\providecommand \href [0]{\begingroup \@sanitize@url \@href}%
\providecommand \@href[1]{\@@startlink{#1}\@@href}%
\providecommand \@@href[1]{\endgroup#1\@@endlink}%
\providecommand \@sanitize@url [0]{\catcode `\\12\catcode `\$12\catcode
  `\&12\catcode `\#12\catcode `\^12\catcode `\_12\catcode `\%12\relax}%
\providecommand \@@startlink[1]{}%
\providecommand \@@endlink[0]{}%
\providecommand \url  [0]{\begingroup\@sanitize@url \@url }%
\providecommand \@url [1]{\endgroup\@href {#1}{\urlprefix }}%
\providecommand \urlprefix  [0]{URL }%
\providecommand \Eprint [0]{\href }%
\providecommand \doibase [0]{http://dx.doi.org/}%
\providecommand \selectlanguage [0]{\@gobble}%
\providecommand \bibinfo  [0]{\@secondoftwo}%
\providecommand \bibfield  [0]{\@secondoftwo}%
\providecommand \translation [1]{[#1]}%
\providecommand \BibitemOpen [0]{}%
\providecommand \bibitemStop [0]{}%
\providecommand \bibitemNoStop [0]{.\EOS\space}%
\providecommand \EOS [0]{\spacefactor3000\relax}%
\providecommand \BibitemShut  [1]{\csname bibitem#1\endcsname}%
\let\auto@bib@innerbib\@empty
%</preamble>
\bibitem [{\citenamefont {Jackson}(1998)}]{Jackson:1998}%
  \BibitemOpen
  \bibfield  {author} {\bibinfo {author} {\bibfnamefont {J.~D.}\ \bibnamefont
  {Jackson}},\ }\href@noop {} {\emph {\bibinfo {title} {Classical
  Electrodynamics}}}\ (\bibinfo  {publisher} {Wiley \& Sons},\ \bibinfo
  {address} {New York},\ \bibinfo {year} {1998})\BibitemShut {NoStop}%
\bibitem [{\citenamefont {Maradudin}\ \emph {et~al.}(2014)\citenamefont
  {Maradudin}, \citenamefont {Sambles},\ and\ \citenamefont
  {Barnes}}]{Maradudin:2014}%
  \BibitemOpen
  \bibfield  {author} {\bibinfo {author} {\bibfnamefont {A.}~\bibnamefont
  {Maradudin}}, \bibinfo {author} {\bibfnamefont {J.~R.}\ \bibnamefont
  {Sambles}}, \ and\ \bibinfo {author} {\bibfnamefont {W.~L.}\ \bibnamefont
  {Barnes}},\ }\href {\doibase 10.1016/B978-0-444-59526-3.00016-1} {\emph
  {\bibinfo {title} {Modern Plasmonics}}}\ (\bibinfo  {publisher}
  {North-Holland},\ \bibinfo {address} {Amsterdam},\ \bibinfo {year}
  {2014})\BibitemShut {NoStop}%
\bibitem [{\citenamefont {Barton}(1979)}]{Barton:1979}%
  \BibitemOpen
  \bibfield  {author} {\bibinfo {author} {\bibfnamefont {G.}~\bibnamefont
  {Barton}},\ }\href {\doibase 10.1088/0034-4885/42/6/001} {\bibfield
  {journal} {\bibinfo  {journal} {Rep. Prog. Phys.}\ }\textbf {\bibinfo
  {volume} {42}},\ \bibinfo {pages} {963} (\bibinfo {year} {1979})}\BibitemShut
  {NoStop}%
\bibitem [{\citenamefont {Pitarke}\ \emph {et~al.}(2007)\citenamefont
  {Pitarke}, \citenamefont {Silkin}, \citenamefont {Chulkov},\ and\
  \citenamefont {Echenique}}]{Pitarke:2007}%
  \BibitemOpen
  \bibfield  {author} {\bibinfo {author} {\bibfnamefont {J.~M.}\ \bibnamefont
  {Pitarke}}, \bibinfo {author} {\bibfnamefont {V.~M.}\ \bibnamefont {Silkin}},
  \bibinfo {author} {\bibfnamefont {E.~V.}\ \bibnamefont {Chulkov}}, \ and\
  \bibinfo {author} {\bibfnamefont {P.~M.}\ \bibnamefont {Echenique}},\ }\href
  {\doibase 10.1088/0034-4885/70/1/R01} {\bibfield  {journal} {\bibinfo
  {journal} {Rep. Prog. Phys.}\ }\textbf {\bibinfo {volume} {70}},\ \bibinfo
  {pages} {1} (\bibinfo {year} {2007})}\BibitemShut {NoStop}%
\bibitem [{\citenamefont {Raza}\ \emph
  {et~al.}(2015{\natexlab{a}})\citenamefont {Raza}, \citenamefont
  {Bozhevolnyi}, \citenamefont {Wubs},\ and\ \citenamefont
  {Mortensen}}]{Raza:2015a}%
  \BibitemOpen
  \bibfield  {author} {\bibinfo {author} {\bibfnamefont {S.}~\bibnamefont
  {Raza}}, \bibinfo {author} {\bibfnamefont {S.~I.}\ \bibnamefont
  {Bozhevolnyi}}, \bibinfo {author} {\bibfnamefont {M.}~\bibnamefont {Wubs}}, \
  and\ \bibinfo {author} {\bibfnamefont {N.~A.}\ \bibnamefont {Mortensen}},\
  }\href {\doibase 10.1088/0953-8984/27/18/183204} {\bibfield  {journal}
  {\bibinfo  {journal} {J. Phys.: Cond. Matter}\ }\textbf {\bibinfo {volume}
  {27}},\ \bibinfo {pages} {183204} (\bibinfo {year}
  {2015}{\natexlab{a}})}\BibitemShut {NoStop}%
\bibitem [{\citenamefont {Stockman}(2011)}]{Stockman:2011}%
  \BibitemOpen
  \bibfield  {author} {\bibinfo {author} {\bibfnamefont {M.~I.}\ \bibnamefont
  {Stockman}},\ }\href {\doibase 10.1364/OE.19.022029} {\bibfield  {journal}
  {\bibinfo  {journal} {Opt. Express}\ }\textbf {\bibinfo {volume} {19}},\
  \bibinfo {pages} {22029} (\bibinfo {year} {2011})}\BibitemShut {NoStop}%
\bibitem [{\citenamefont {Fern{\'a}ndez-Dom{\'i}nguez}\ \emph
  {et~al.}(2017)\citenamefont {Fern{\'a}ndez-Dom{\'i}nguez}, \citenamefont
  {Garc{\'i}a-Vidal},\ and\ \citenamefont
  {Mart{\'i}n-Moreno}}]{Fernandez-Dominguez:2017}%
  \BibitemOpen
  \bibfield  {author} {\bibinfo {author} {\bibfnamefont {A.~I.}\ \bibnamefont
  {Fern{\'a}ndez-Dom{\'i}nguez}}, \bibinfo {author} {\bibfnamefont {F.~J.}\
  \bibnamefont {Garc{\'i}a-Vidal}}, \ and\ \bibinfo {author} {\bibfnamefont
  {L.}~\bibnamefont {Mart{\'i}n-Moreno}},\ }\href {\doibase
  10.1038/nphoton.2016.258} {\bibfield  {journal} {\bibinfo  {journal} {Nat.
  Photon.}\ }\textbf {\bibinfo {volume} {11}},\ \bibinfo {pages} {8} (\bibinfo
  {year} {2017})}\BibitemShut {NoStop}%
\bibitem [{\citenamefont {Keller}\ \emph {et~al.}(1993)\citenamefont {Keller},
  \citenamefont {Xiao},\ and\ \citenamefont {Bozhevolnyi}}]{Keller:1993}%
  \BibitemOpen
  \bibfield  {author} {\bibinfo {author} {\bibfnamefont {O.}~\bibnamefont
  {Keller}}, \bibinfo {author} {\bibfnamefont {M.}~\bibnamefont {Xiao}}, \ and\
  \bibinfo {author} {\bibfnamefont {S.~I.}\ \bibnamefont {Bozhevolnyi}},\
  }\href {\doibase 10.1016/0030-4018(93)90389-M} {\bibfield  {journal}
  {\bibinfo  {journal} {Opt. Commun.}\ }\textbf {\bibinfo {volume} {102}},\
  \bibinfo {pages} {238} (\bibinfo {year} {1993})}\BibitemShut {NoStop}%
\bibitem [{\citenamefont {{\"O}zt{\"u}rk}\ \emph {et~al.}(2011)\citenamefont
  {{\"O}zt{\"u}rk}, \citenamefont {Xiao}, \citenamefont {Yan}, \citenamefont
  {Wubs}, \citenamefont {Jauho},\ and\ \citenamefont
  {Mortensen}}]{Ozturk:2011}%
  \BibitemOpen
  \bibfield  {author} {\bibinfo {author} {\bibfnamefont {Z.~F.}\ \bibnamefont
  {{\"O}zt{\"u}rk}}, \bibinfo {author} {\bibfnamefont {S.}~\bibnamefont
  {Xiao}}, \bibinfo {author} {\bibfnamefont {M.}~\bibnamefont {Yan}}, \bibinfo
  {author} {\bibfnamefont {M.}~\bibnamefont {Wubs}}, \bibinfo {author}
  {\bibfnamefont {A.-P.}\ \bibnamefont {Jauho}}, \ and\ \bibinfo {author}
  {\bibfnamefont {N.~A.}\ \bibnamefont {Mortensen}},\ }\href {\doibase
  10.1117/1.3574159} {\bibfield  {journal} {\bibinfo  {journal} {J.
  Nanophoton.}\ }\textbf {\bibinfo {volume} {5}},\ \bibinfo {pages} {051602}
  (\bibinfo {year} {2011})}\BibitemShut {NoStop}%
\bibitem [{\citenamefont {David}\ and\ \citenamefont {Garc\'{i}a~de
  Abajo}(2014)}]{David:2014}%
  \BibitemOpen
  \bibfield  {author} {\bibinfo {author} {\bibfnamefont {C.}~\bibnamefont
  {David}}\ and\ \bibinfo {author} {\bibfnamefont {F.~J.}\ \bibnamefont
  {Garc\'{i}a~de Abajo}},\ }\href {\doibase 10.1021/nn5038527} {\bibfield
  {journal} {\bibinfo  {journal} {ACS Nano}\ }\textbf {\bibinfo {volume} {8}},\
  \bibinfo {pages} {9558} (\bibinfo {year} {2014})}\BibitemShut {NoStop}%
\bibitem [{\citenamefont {{Garc\'{i}a de Abajo}}(2008)}]{GarciadeAbajo:2008}%
  \BibitemOpen
  \bibfield  {author} {\bibinfo {author} {\bibfnamefont {F.~J.}\ \bibnamefont
  {{Garc\'{i}a de Abajo}}},\ }\href {\doibase 10.1021/jp807345h} {\bibfield
  {journal} {\bibinfo  {journal} {J. Phys. Chem. C}\ }\textbf {\bibinfo
  {volume} {112}},\ \bibinfo {pages} {17983} (\bibinfo {year}
  {2008})}\BibitemShut {NoStop}%
\bibitem [{\citenamefont {Mortensen}\ \emph {et~al.}(2014)\citenamefont
  {Mortensen}, \citenamefont {Raza}, \citenamefont {Wubs}, \citenamefont
  {S{\o}ndergaard},\ and\ \citenamefont {Bozhevolnyi}}]{Mortensen:2014}%
  \BibitemOpen
  \bibfield  {author} {\bibinfo {author} {\bibfnamefont {N.~A.}\ \bibnamefont
  {Mortensen}}, \bibinfo {author} {\bibfnamefont {S.}~\bibnamefont {Raza}},
  \bibinfo {author} {\bibfnamefont {M.}~\bibnamefont {Wubs}}, \bibinfo {author}
  {\bibfnamefont {T.}~\bibnamefont {S{\o}ndergaard}}, \ and\ \bibinfo {author}
  {\bibfnamefont {S.~I.}\ \bibnamefont {Bozhevolnyi}},\ }\href {\doibase
  10.1038/ncomms4809} {\bibfield  {journal} {\bibinfo  {journal} {Nat.
  Commun.}\ }\textbf {\bibinfo {volume} {5}},\ \bibinfo {pages} {3809}
  (\bibinfo {year} {2014})}\BibitemShut {NoStop}%
\bibitem [{\citenamefont {Toscano}\ \emph {et~al.}(2015)\citenamefont
  {Toscano}, \citenamefont {Straubel}, \citenamefont {Kwiatkowski},
  \citenamefont {Rockstuhl}, \citenamefont {Evers}, \citenamefont {Xu},
  \citenamefont {Mortensen},\ and\ \citenamefont {Wubs}}]{Toscano:2015}%
  \BibitemOpen
  \bibfield  {author} {\bibinfo {author} {\bibfnamefont {G.}~\bibnamefont
  {Toscano}}, \bibinfo {author} {\bibfnamefont {J.}~\bibnamefont {Straubel}},
  \bibinfo {author} {\bibfnamefont {A.}~\bibnamefont {Kwiatkowski}}, \bibinfo
  {author} {\bibfnamefont {C.}~\bibnamefont {Rockstuhl}}, \bibinfo {author}
  {\bibfnamefont {F.}~\bibnamefont {Evers}}, \bibinfo {author} {\bibfnamefont
  {H.}~\bibnamefont {Xu}}, \bibinfo {author} {\bibfnamefont {N.~A.}\
  \bibnamefont {Mortensen}}, \ and\ \bibinfo {author} {\bibfnamefont
  {M.}~\bibnamefont {Wubs}},\ }\href {\doibase 10.1038/ncomms8132} {\bibfield
  {journal} {\bibinfo  {journal} {Nat. Commun.}\ }\textbf {\bibinfo {volume}
  {6}},\ \bibinfo {pages} {7132} (\bibinfo {year} {2015})}\BibitemShut
  {NoStop}%
\bibitem [{\citenamefont {Varas}\ \emph {et~al.}(2016)\citenamefont {Varas},
  \citenamefont {Garc{\'i}a-Gonz{\'a}lez}, \citenamefont {Feist}, \citenamefont
  {Garc{\'i}a-Vidal},\ and\ \citenamefont {Rubio}}]{Varas:2016}%
  \BibitemOpen
  \bibfield  {author} {\bibinfo {author} {\bibfnamefont {A.}~\bibnamefont
  {Varas}}, \bibinfo {author} {\bibfnamefont {P.}~\bibnamefont
  {Garc{\'i}a-Gonz{\'a}lez}}, \bibinfo {author} {\bibfnamefont
  {J.}~\bibnamefont {Feist}}, \bibinfo {author} {\bibfnamefont {F.~J.}\
  \bibnamefont {Garc{\'i}a-Vidal}}, \ and\ \bibinfo {author} {\bibfnamefont
  {A.}~\bibnamefont {Rubio}},\ }\href {\doibase 10.1515/nanoph-2015-0141}
  {\bibfield  {journal} {\bibinfo  {journal} {Nanophotonics}\ }\textbf
  {\bibinfo {volume} {5}},\ \bibinfo {pages} {409} (\bibinfo {year}
  {2016})}\BibitemShut {NoStop}%
\bibitem [{\citenamefont {Zhu}\ \emph {et~al.}(2016)\citenamefont {Zhu},
  \citenamefont {Esteban}, \citenamefont {Borisov}, \citenamefont {Baumberg},
  \citenamefont {Nordlander}, \citenamefont {Lezec}, \citenamefont {Aizpurua},\
  and\ \citenamefont {Crozier}}]{Zhu:2016}%
  \BibitemOpen
  \bibfield  {author} {\bibinfo {author} {\bibfnamefont {W.}~\bibnamefont
  {Zhu}}, \bibinfo {author} {\bibfnamefont {R.}~\bibnamefont {Esteban}},
  \bibinfo {author} {\bibfnamefont {A.~G.}\ \bibnamefont {Borisov}}, \bibinfo
  {author} {\bibfnamefont {J.~J.}\ \bibnamefont {Baumberg}}, \bibinfo {author}
  {\bibfnamefont {P.}~\bibnamefont {Nordlander}}, \bibinfo {author}
  {\bibfnamefont {H.~J.}\ \bibnamefont {Lezec}}, \bibinfo {author}
  {\bibfnamefont {J.}~\bibnamefont {Aizpurua}}, \ and\ \bibinfo {author}
  {\bibfnamefont {K.~B.}\ \bibnamefont {Crozier}},\ }\href {\doibase
  10.1038/ncomms11495} {\bibfield  {journal} {\bibinfo  {journal} {Nat.
  Commun.}\ }\textbf {\bibinfo {volume} {7}},\ \bibinfo {pages} {11495}
  (\bibinfo {year} {2016})}\BibitemShut {NoStop}%
\bibitem [{\citenamefont {Zuloaga}\ \emph {et~al.}(2009)\citenamefont
  {Zuloaga}, \citenamefont {Prodan},\ and\ \citenamefont
  {Nordlander}}]{Zuloaga:2009}%
  \BibitemOpen
  \bibfield  {author} {\bibinfo {author} {\bibfnamefont {J.}~\bibnamefont
  {Zuloaga}}, \bibinfo {author} {\bibfnamefont {E.}~\bibnamefont {Prodan}}, \
  and\ \bibinfo {author} {\bibfnamefont {P.}~\bibnamefont {Nordlander}},\
  }\href {\doibase 10.1021/nl803811g} {\bibfield  {journal} {\bibinfo
  {journal} {Nano Lett.}\ }\textbf {\bibinfo {volume} {9}},\ \bibinfo {pages}
  {887} (\bibinfo {year} {2009})}\BibitemShut {NoStop}%
\bibitem [{\citenamefont {Zuloaga}\ \emph {et~al.}(2010)\citenamefont
  {Zuloaga}, \citenamefont {Prodan},\ and\ \citenamefont
  {Nordlander}}]{Zuloaga:2010}%
  \BibitemOpen
  \bibfield  {author} {\bibinfo {author} {\bibfnamefont {J.}~\bibnamefont
  {Zuloaga}}, \bibinfo {author} {\bibfnamefont {E.}~\bibnamefont {Prodan}}, \
  and\ \bibinfo {author} {\bibfnamefont {P.}~\bibnamefont {Nordlander}},\
  }\href {\doibase 10.1021/nn101589n} {\bibfield  {journal} {\bibinfo
  {journal} {ACS Nano}\ }\textbf {\bibinfo {volume} {4}},\ \bibinfo {pages}
  {5269} (\bibinfo {year} {2010})}\BibitemShut {NoStop}%
\bibitem [{\citenamefont {Teperik}\ \emph {et~al.}(2013)\citenamefont
  {Teperik}, \citenamefont {Nordlander}, \citenamefont {Aizpurua},\ and\
  \citenamefont {Borisov}}]{Teperik:2013}%
  \BibitemOpen
  \bibfield  {author} {\bibinfo {author} {\bibfnamefont {T.~V.}\ \bibnamefont
  {Teperik}}, \bibinfo {author} {\bibfnamefont {P.}~\bibnamefont {Nordlander}},
  \bibinfo {author} {\bibfnamefont {J.}~\bibnamefont {Aizpurua}}, \ and\
  \bibinfo {author} {\bibfnamefont {A.~G.}\ \bibnamefont {Borisov}},\ }\href
  {\doibase 10.1103/PhysRevLett.110.263901} {\bibfield  {journal} {\bibinfo
  {journal} {Phys. Rev. Lett.}\ }\textbf {\bibinfo {volume} {110}},\ \bibinfo
  {pages} {263901} (\bibinfo {year} {2013})}\BibitemShut {NoStop}%
\bibitem [{\citenamefont {Andersen}\ \emph {et~al.}(2013)\citenamefont
  {Andersen}, \citenamefont {Jensen}, \citenamefont {Mortensen},\ and\
  \citenamefont {Thygesen}}]{Andersen:2013}%
  \BibitemOpen
  \bibfield  {author} {\bibinfo {author} {\bibfnamefont {K.}~\bibnamefont
  {Andersen}}, \bibinfo {author} {\bibfnamefont {K.~L.}\ \bibnamefont
  {Jensen}}, \bibinfo {author} {\bibfnamefont {N.~A.}\ \bibnamefont
  {Mortensen}}, \ and\ \bibinfo {author} {\bibfnamefont {K.~S.}\ \bibnamefont
  {Thygesen}},\ }\href {\doibase 10.1103/PhysRevB.87.235433} {\bibfield
  {journal} {\bibinfo  {journal} {Phys. Rev. B}\ }\textbf {\bibinfo {volume}
  {87}},\ \bibinfo {pages} {235433} (\bibinfo {year} {2013})}\BibitemShut
  {NoStop}%
\bibitem [{\citenamefont {Sinha-Roy}\ \emph {et~al.}(2017)\citenamefont
  {Sinha-Roy}, \citenamefont {Garc{\'i}a-Gonz{\'a}lez}, \citenamefont
  {Weissker}, \citenamefont {Rabilloud},\ and\ \citenamefont
  {Fern{\'a}ndez-Dom{\'i}nguez}}]{Sinha-Roy:2017}%
  \BibitemOpen
  \bibfield  {author} {\bibinfo {author} {\bibfnamefont {R.}~\bibnamefont
  {Sinha-Roy}}, \bibinfo {author} {\bibfnamefont {P.}~\bibnamefont
  {Garc{\'i}a-Gonz{\'a}lez}}, \bibinfo {author} {\bibfnamefont {H.-C.}\
  \bibnamefont {Weissker}}, \bibinfo {author} {\bibfnamefont {F.}~\bibnamefont
  {Rabilloud}}, \ and\ \bibinfo {author} {\bibfnamefont {A.~I.}\ \bibnamefont
  {Fern{\'a}ndez-Dom{\'i}nguez}},\ }\href {\doibase
  10.1021/acsphotonics.7b00254} {\bibfield  {journal} {\bibinfo  {journal} {ACS
  Photonics}\ }\textbf {\bibinfo {volume} {4}},\ \bibinfo {pages} {1484}
  (\bibinfo {year} {2017})}\BibitemShut {NoStop}%
\bibitem [{\citenamefont {Feibelman}(1982)}]{Feibelman:1982}%
  \BibitemOpen
  \bibfield  {author} {\bibinfo {author} {\bibfnamefont {P.~J.}\ \bibnamefont
  {Feibelman}},\ }\href {\doibase 10.1016/0079-6816(82)90001-6} {\bibfield
  {journal} {\bibinfo  {journal} {Prog. Surf. Sci.}\ }\textbf {\bibinfo
  {volume} {12}},\ \bibinfo {pages} {287} (\bibinfo {year} {1982})}\BibitemShut
  {NoStop}%
\bibitem [{\citenamefont {Liebsch}(1997)}]{Liebsch:1997}%
  \BibitemOpen
  \bibfield  {author} {\bibinfo {author} {\bibfnamefont {A.}~\bibnamefont
  {Liebsch}},\ }\href {\doibase 10.1007/978-1-4757-5107-9} {\emph {\bibinfo
  {title} {Electronic Excitations at Metal Surfaces}}}\ (\bibinfo  {publisher}
  {Springer},\ \bibinfo {address} {New York},\ \bibinfo {year}
  {1997})\BibitemShut {NoStop}%
\bibitem [{\citenamefont {Deng}(2020)}]{Deng:2020}%
  \BibitemOpen
  \bibfield  {author} {\bibinfo {author} {\bibfnamefont {H.-Y.}\ \bibnamefont
  {Deng}},\ }\href {\doibase 10.1016/j.aop.2020.168204} {\bibfield  {journal}
  {\bibinfo  {journal} {Ann. Phys.}\ }\textbf {\bibinfo {volume} {418}},\
  \bibinfo {pages} {168204} (\bibinfo {year} {2020})}\BibitemShut {NoStop}%
\bibitem [{\citenamefont {Bozhevolnyi}\ and\ \citenamefont
  {Mortensen}(2017)}]{Bozhevolnyi:2017a}%
  \BibitemOpen
  \bibfield  {author} {\bibinfo {author} {\bibfnamefont {S.~I.}\ \bibnamefont
  {Bozhevolnyi}}\ and\ \bibinfo {author} {\bibfnamefont {N.~A.}\ \bibnamefont
  {Mortensen}},\ }\href {\doibase 10.1515/nanoph-2016-0179} {\bibfield
  {journal} {\bibinfo  {journal} {Nanophotonics}\ }\textbf {\bibinfo {volume}
  {6}},\ \bibinfo {pages} {1185} (\bibinfo {year} {2017})}\BibitemShut
  {NoStop}%
\bibitem [{\citenamefont {Fern{\'a}ndez-Dom{\'i}nguez}\ \emph
  {et~al.}(2018)\citenamefont {Fern{\'a}ndez-Dom{\'i}nguez}, \citenamefont
  {Bozhevolnyi},\ and\ \citenamefont {Mortensen}}]{Fernandez-Dominguez:2018}%
  \BibitemOpen
  \bibfield  {author} {\bibinfo {author} {\bibfnamefont {A.~I.}\ \bibnamefont
  {Fern{\'a}ndez-Dom{\'i}nguez}}, \bibinfo {author} {\bibfnamefont {S.~I.}\
  \bibnamefont {Bozhevolnyi}}, \ and\ \bibinfo {author} {\bibfnamefont {N.~A.}\
  \bibnamefont {Mortensen}},\ }\href {\doibase 10.1021/acsphotonics.8b00852}
  {\bibfield  {journal} {\bibinfo  {journal} {ACS Photonics}\ }\textbf
  {\bibinfo {volume} {5}},\ \bibinfo {pages} {3447} (\bibinfo {year}
  {2018})}\BibitemShut {NoStop}%
\bibitem [{\citenamefont {Gon\c{c}alves}(2020)}]{Goncalves_SpringerTheses}%
  \BibitemOpen
  \bibfield  {author} {\bibinfo {author} {\bibfnamefont {P.~A.~D.}\
  \bibnamefont {Gon\c{c}alves}},\ }\href {\doibase 10.1007/978-3-030-38291-9}
  {\emph {\bibinfo {title} {Plasmonics and Light--Matter Interactions in
  Two-Dimensional Materials and in Metal Nanostructures: Classical and Quantum
  Considerations}}}\ (\bibinfo  {publisher} {Springer Nature},\ \bibinfo {year}
  {2020})\BibitemShut {NoStop}%
\bibitem [{\citenamefont {Gon\c{c}alves}\ \emph
  {et~al.}(2020{\natexlab{a}})\citenamefont {Gon\c{c}alves}, \citenamefont
  {Christensen}, \citenamefont {Rivera}, \citenamefont {Jauho}, \citenamefont
  {Mortensen},\ and\ \citenamefont {Solja\v{c}i\'{c}}}]{Goncalves:2020}%
  \BibitemOpen
  \bibfield  {author} {\bibinfo {author} {\bibfnamefont {P.~A.~D.}\
  \bibnamefont {Gon\c{c}alves}}, \bibinfo {author} {\bibfnamefont
  {T.}~\bibnamefont {Christensen}}, \bibinfo {author} {\bibfnamefont
  {N.}~\bibnamefont {Rivera}}, \bibinfo {author} {\bibfnamefont {A.-P.}\
  \bibnamefont {Jauho}}, \bibinfo {author} {\bibfnamefont {N.~A.}\ \bibnamefont
  {Mortensen}}, \ and\ \bibinfo {author} {\bibfnamefont {M.}~\bibnamefont
  {Solja\v{c}i\'{c}}},\ }\href {\doibase 10.1038/s41467-019-13820-z} {\bibfield
   {journal} {\bibinfo  {journal} {Nat. Commun.}\ }\textbf {\bibinfo {volume}
  {11}},\ \bibinfo {pages} {366} (\bibinfo {year}
  {2020}{\natexlab{a}})}\BibitemShut {NoStop}%
\bibitem [{\citenamefont {Tame}\ \emph {et~al.}(2013)\citenamefont {Tame},
  \citenamefont {McEnery}, \citenamefont {\c{S}. K.~{\"O}zdemir}, \citenamefont
  {Lee}, \citenamefont {Maier},\ and\ \citenamefont {Kim}}]{Tame:2013}%
  \BibitemOpen
  \bibfield  {author} {\bibinfo {author} {\bibfnamefont {M.~S.}\ \bibnamefont
  {Tame}}, \bibinfo {author} {\bibfnamefont {K.~R.}\ \bibnamefont {McEnery}},
  \bibinfo {author} {\bibnamefont {\c{S}. K.~{\"O}zdemir}}, \bibinfo {author}
  {\bibfnamefont {J.}~\bibnamefont {Lee}}, \bibinfo {author} {\bibfnamefont
  {S.~A.}\ \bibnamefont {Maier}}, \ and\ \bibinfo {author} {\bibfnamefont
  {M.~S.}\ \bibnamefont {Kim}},\ }\href {\doibase 10.1038/nphys2615} {\bibfield
   {journal} {\bibinfo  {journal} {Nat. Phys.}\ }\textbf {\bibinfo {volume}
  {9}},\ \bibinfo {pages} {329} (\bibinfo {year} {2013})}\BibitemShut {NoStop}%
\bibitem [{\citenamefont {Fitzgerald}\ \emph {et~al.}(2016)\citenamefont
  {Fitzgerald}, \citenamefont {Narang}, \citenamefont {Craster}, \citenamefont
  {Maier},\ and\ \citenamefont {Giannini}}]{Fitzgerald:2016}%
  \BibitemOpen
  \bibfield  {author} {\bibinfo {author} {\bibfnamefont {J.~M.}\ \bibnamefont
  {Fitzgerald}}, \bibinfo {author} {\bibfnamefont {P.}~\bibnamefont {Narang}},
  \bibinfo {author} {\bibfnamefont {R.~V.}\ \bibnamefont {Craster}}, \bibinfo
  {author} {\bibfnamefont {S.~A.}\ \bibnamefont {Maier}}, \ and\ \bibinfo
  {author} {\bibfnamefont {V.}~\bibnamefont {Giannini}},\ }\href {\doibase
  10.1109/JPROC.2016.2584860} {\bibfield  {journal} {\bibinfo  {journal} {Proc.
  IEEE}\ }\textbf {\bibinfo {volume} {104}},\ \bibinfo {pages} {2307} (\bibinfo
  {year} {2016})}\BibitemShut {NoStop}%
\bibitem [{\citenamefont {Zhou}\ \emph {et~al.}(2019)\citenamefont {Zhou},
  \citenamefont {Liu}, \citenamefont {Bao}, \citenamefont {Wu}, \citenamefont
  {Png}, \citenamefont {Wang},\ and\ \citenamefont {Qiu}}]{Zhou:2019}%
  \BibitemOpen
  \bibfield  {author} {\bibinfo {author} {\bibfnamefont {Z.-K.}\ \bibnamefont
  {Zhou}}, \bibinfo {author} {\bibfnamefont {J.}~\bibnamefont {Liu}}, \bibinfo
  {author} {\bibfnamefont {Y.}~\bibnamefont {Bao}}, \bibinfo {author}
  {\bibfnamefont {L.}~\bibnamefont {Wu}}, \bibinfo {author} {\bibfnamefont
  {C.~E.}\ \bibnamefont {Png}}, \bibinfo {author} {\bibfnamefont {X.-H.}\
  \bibnamefont {Wang}}, \ and\ \bibinfo {author} {\bibfnamefont {C.-W.}\
  \bibnamefont {Qiu}},\ }\href {\doibase 10.1016/j.pquantelec.2019.04.002}
  {\bibfield  {journal} {\bibinfo  {journal} {Prog. Quantum Electron.}\
  }\textbf {\bibinfo {volume} {65}},\ \bibinfo {pages} {1} (\bibinfo {year}
  {2019})}\BibitemShut {NoStop}%
\bibitem [{\citenamefont {Gon\c{c}alves}\ \emph {et~al.}()\citenamefont
  {Gon\c{c}alves}, \citenamefont {Christensen}, \citenamefont {Peres},
  \citenamefont {Jauho}, \citenamefont {Epstein}, \citenamefont {Koppens},
  \citenamefont {Solja\v{c}i\'{c}},\ and\ \citenamefont
  {Mortensen}}]{Goncalves:2020a}%
  \BibitemOpen
  \bibfield  {author} {\bibinfo {author} {\bibfnamefont {P.~A.~D.}\
  \bibnamefont {Gon\c{c}alves}}, \bibinfo {author} {\bibfnamefont
  {T.}~\bibnamefont {Christensen}}, \bibinfo {author} {\bibfnamefont
  {N.~M.~R.}\ \bibnamefont {Peres}}, \bibinfo {author} {\bibfnamefont {A.-P.}\
  \bibnamefont {Jauho}}, \bibinfo {author} {\bibfnamefont {I.}~\bibnamefont
  {Epstein}}, \bibinfo {author} {\bibfnamefont {F.~H.~L.}\ \bibnamefont
  {Koppens}}, \bibinfo {author} {\bibfnamefont {M.}~\bibnamefont
  {Solja\v{c}i\'{c}}}, \ and\ \bibinfo {author} {\bibfnamefont {N.~A.}\
  \bibnamefont {Mortensen}},\ }\href {https://arxiv.org/abs/2008.07613}
  {\bibinfo  {journal} {arXiv:2008.07613}\ }\BibitemShut {NoStop}%
\bibitem [{\citenamefont {Apell}\ and\ \citenamefont
  {Penn}(1983)}]{Apell:1983}%
  \BibitemOpen
\bibfield  {journal} {  }\bibfield  {author} {\bibinfo {author} {\bibfnamefont
  {P.}~\bibnamefont {Apell}}\ and\ \bibinfo {author} {\bibfnamefont {D.~R.}\
  \bibnamefont {Penn}},\ }\href {\doibase 10.1103/PhysRevLett.50.1316}
  {\bibfield  {journal} {\bibinfo  {journal} {Phys. Rev. Lett.}\ }\textbf
  {\bibinfo {volume} {50}},\ \bibinfo {pages} {1316} (\bibinfo {year}
  {1983})}\BibitemShut {NoStop}%
\bibitem [{\citenamefont {Yan}\ \emph {et~al.}(2015)\citenamefont {Yan},
  \citenamefont {Wubs},\ and\ \citenamefont {Mortensen}}]{Yan:2015}%
  \BibitemOpen
  \bibfield  {author} {\bibinfo {author} {\bibfnamefont {W.}~\bibnamefont
  {Yan}}, \bibinfo {author} {\bibfnamefont {M.}~\bibnamefont {Wubs}}, \ and\
  \bibinfo {author} {\bibfnamefont {N.~A.}\ \bibnamefont {Mortensen}},\ }\href
  {\doibase 10.1103/PhysRevLett.115.137403} {\bibfield  {journal} {\bibinfo
  {journal} {Phys. Rev. Lett.}\ }\textbf {\bibinfo {volume} {115}},\ \bibinfo
  {pages} {137403} (\bibinfo {year} {2015})}\BibitemShut {NoStop}%
\bibitem [{\citenamefont {Christensen}\ \emph {et~al.}(2017)\citenamefont
  {Christensen}, \citenamefont {Yan}, \citenamefont {Jauho}, \citenamefont
  {Solja\v{c}i\'{c}},\ and\ \citenamefont {Mortensen}}]{Christensen:2017}%
  \BibitemOpen
  \bibfield  {author} {\bibinfo {author} {\bibfnamefont {T.}~\bibnamefont
  {Christensen}}, \bibinfo {author} {\bibfnamefont {W.}~\bibnamefont {Yan}},
  \bibinfo {author} {\bibfnamefont {A.-P.}\ \bibnamefont {Jauho}}, \bibinfo
  {author} {\bibfnamefont {M.}~\bibnamefont {Solja\v{c}i\'{c}}}, \ and\
  \bibinfo {author} {\bibfnamefont {N.~A.}\ \bibnamefont {Mortensen}},\ }\href
  {\doibase 10.1103/PhysRevLett.118.157402} {\bibfield  {journal} {\bibinfo
  {journal} {Phys. Rev. Lett.}\ }\textbf {\bibinfo {volume} {118}},\ \bibinfo
  {pages} {157402} (\bibinfo {year} {2017})}\BibitemShut {NoStop}%
\bibitem [{\citenamefont {Yang}\ \emph {et~al.}(2019)\citenamefont {Yang},
  \citenamefont {Zhu}, \citenamefont {Yan}, \citenamefont {Agarwal},
  \citenamefont {Zheng}, \citenamefont {Joannopoulos}, \citenamefont {Lalanne},
  \citenamefont {Christensen}, \citenamefont {Berggren},\ and\ \citenamefont
  {Solja\v{c}i\'{c}}}]{Yang:2019}%
  \BibitemOpen
  \bibfield  {author} {\bibinfo {author} {\bibfnamefont {Y.}~\bibnamefont
  {Yang}}, \bibinfo {author} {\bibfnamefont {D.}~\bibnamefont {Zhu}}, \bibinfo
  {author} {\bibfnamefont {W.}~\bibnamefont {Yan}}, \bibinfo {author}
  {\bibfnamefont {A.}~\bibnamefont {Agarwal}}, \bibinfo {author} {\bibfnamefont
  {M.}~\bibnamefont {Zheng}}, \bibinfo {author} {\bibfnamefont {J.~D.}\
  \bibnamefont {Joannopoulos}}, \bibinfo {author} {\bibfnamefont
  {P.}~\bibnamefont {Lalanne}}, \bibinfo {author} {\bibfnamefont
  {T.}~\bibnamefont {Christensen}}, \bibinfo {author} {\bibfnamefont {K.~K.}\
  \bibnamefont {Berggren}}, \ and\ \bibinfo {author} {\bibfnamefont
  {M.}~\bibnamefont {Solja\v{c}i\'{c}}},\ }\href {\doibase
  10.1038/s41586-019-1803-1} {\bibfield  {journal} {\bibinfo  {journal}
  {Nature}\ }\textbf {\bibinfo {volume} {576}},\ \bibinfo {pages} {248}
  (\bibinfo {year} {2019})}\BibitemShut {NoStop}%
\bibitem [{\citenamefont {Echarri}\ \emph {et~al.}()\citenamefont {Echarri},
  \citenamefont {Gonçalves}, \citenamefont {Tserkezis}, \citenamefont
  {Garc{\'i}a~de Abajo}, \citenamefont {Mortensen},\ and\ \citenamefont
  {Cox}}]{Echarri:2020}%
  \BibitemOpen
  \bibfield  {author} {\bibinfo {author} {\bibfnamefont {A.~R.}\ \bibnamefont
  {Echarri}}, \bibinfo {author} {\bibfnamefont {P.~A.~D.}\ \bibnamefont
  {Gonçalves}}, \bibinfo {author} {\bibfnamefont {C.}~\bibnamefont
  {Tserkezis}}, \bibinfo {author} {\bibfnamefont {F.~J.}\ \bibnamefont
  {Garc{\'i}a~de Abajo}}, \bibinfo {author} {\bibfnamefont {N.~A.}\
  \bibnamefont {Mortensen}}, \ and\ \bibinfo {author} {\bibfnamefont {J.~D.}\
  \bibnamefont {Cox}},\ }\href {https://arxiv.org/abs/2009.10821} {\bibinfo
  {journal} {arXiv:2009.10821}\ }\BibitemShut {NoStop}%
\bibitem [{\citenamefont {Langreth}\ and\ \citenamefont
  {Suhl}(1984)}]{Langreth:1984}%
  \BibitemOpen
\bibfield  {journal} {  }\bibfield  {author} {\bibinfo {author} {\bibfnamefont
  {D.}~\bibnamefont {Langreth}}\ and\ \bibinfo {author} {\bibfnamefont
  {H.}~\bibnamefont {Suhl}},\ }\href {\doibase
  10.1016/B978-0-124-36560-5.X5001-2} {\emph {\bibinfo {title} {Many-Body
  Phenomena at Surfaces}}}\ (\bibinfo  {publisher} {Academic Press},\ \bibinfo
  {address} {Orlando},\ \bibinfo {year} {1984})\BibitemShut {NoStop}%
\bibitem [{\citenamefont {Feibelman}(1975)}]{Feibelman:1975}%
  \BibitemOpen
  \bibfield  {author} {\bibinfo {author} {\bibfnamefont {P.~J.}\ \bibnamefont
  {Feibelman}},\ }\href {\doibase 10.1103/PhysRevB.12.1319} {\bibfield
  {journal} {\bibinfo  {journal} {Phys. Rev. B}\ }\textbf {\bibinfo {volume}
  {12}},\ \bibinfo {pages} {1319} (\bibinfo {year} {1975})}\BibitemShut
  {NoStop}%
\bibitem [{\citenamefont {Svendsen}\ \emph {et~al.}(2020)\citenamefont
  {Svendsen}, \citenamefont {Wolff}, \citenamefont {Jauho}, \citenamefont
  {Mortensen},\ and\ \citenamefont {Tserkezis}}]{Svendsen:2020}%
  \BibitemOpen
  \bibfield  {author} {\bibinfo {author} {\bibfnamefont {M.~K.}\ \bibnamefont
  {Svendsen}}, \bibinfo {author} {\bibfnamefont {C.}~\bibnamefont {Wolff}},
  \bibinfo {author} {\bibfnamefont {A.-P.}\ \bibnamefont {Jauho}}, \bibinfo
  {author} {\bibfnamefont {N.~A.}\ \bibnamefont {Mortensen}}, \ and\ \bibinfo
  {author} {\bibfnamefont {C.}~\bibnamefont {Tserkezis}},\ }\href {\doibase
  10.1088/1361-648X/ab977d} {\bibfield  {journal} {\bibinfo  {journal} {J.
  Phys.: Cond. Matter}\ }\textbf {\bibinfo {volume} {32}},\ \bibinfo {pages}
  {395702} (\bibinfo {year} {2020})}\BibitemShut {NoStop}%
\bibitem [{\citenamefont {Keller}(1986)}]{Keller:1986}%
  \BibitemOpen
  \bibfield  {author} {\bibinfo {author} {\bibfnamefont {O.}~\bibnamefont
  {Keller}},\ }\href {\doibase 10.1103/PhysRevB.33.990} {\bibfield  {journal}
  {\bibinfo  {journal} {Phys. Rev. B}\ }\textbf {\bibinfo {volume} {33}},\
  \bibinfo {pages} {990} (\bibinfo {year} {1986})}\BibitemShut {NoStop}%
\bibitem [{\citenamefont {Ichikawa}(2011)}]{Ichikawa:2011}%
  \BibitemOpen
  \bibfield  {author} {\bibinfo {author} {\bibfnamefont {M.}~\bibnamefont
  {Ichikawa}},\ }\href {\doibase 10.1143/JPSJ.80.044606} {\bibfield  {journal}
  {\bibinfo  {journal} {J. Phys. Soc. Jpn.}\ }\textbf {\bibinfo {volume}
  {80}},\ \bibinfo {pages} {044606} (\bibinfo {year} {2011})}\BibitemShut
  {NoStop}%
\bibitem [{\citenamefont {Ichikawa}(2018)}]{Ichikawa:2018}%
  \BibitemOpen
  \bibfield  {author} {\bibinfo {author} {\bibfnamefont {M.}~\bibnamefont
  {Ichikawa}},\ }\href {\doibase 10.1380/ejssnt.2018.329} {\bibfield  {journal}
  {\bibinfo  {journal} {e-J. Surf. Sci. Nanotec.}\ }\textbf {\bibinfo {volume}
  {16}},\ \bibinfo {pages} {329} (\bibinfo {year} {2018})}\BibitemShut
  {NoStop}%
\bibitem [{\citenamefont {Friedel}(1952)}]{Friedel:1952}%
  \BibitemOpen
  \bibfield  {author} {\bibinfo {author} {\bibfnamefont {J.}~\bibnamefont
  {Friedel}},\ }\href {\doibase 10.1080/14786440208561086} {\bibfield
  {journal} {\bibinfo  {journal} {Philos. Mag.}\ }\textbf {\bibinfo {volume}
  {43}},\ \bibinfo {pages} {153} (\bibinfo {year} {1952})}\BibitemShut
  {NoStop}%
\bibitem [{\citenamefont {Lang}\ and\ \citenamefont {Kohn}(1970)}]{Lang:1970}%
  \BibitemOpen
  \bibfield  {author} {\bibinfo {author} {\bibfnamefont {N.~D.}\ \bibnamefont
  {Lang}}\ and\ \bibinfo {author} {\bibfnamefont {W.}~\bibnamefont {Kohn}},\
  }\href {\doibase 10.1103/PhysRevB.1.4555} {\bibfield  {journal} {\bibinfo
  {journal} {Phys. Rev. B}\ }\textbf {\bibinfo {volume} {1}},\ \bibinfo {pages}
  {4555} (\bibinfo {year} {1970})}\BibitemShut {NoStop}%
\bibitem [{\citenamefont {Lindhard}(1954)}]{Lindhard:1954}%
  \BibitemOpen
  \bibfield  {author} {\bibinfo {author} {\bibfnamefont {J.}~\bibnamefont
  {Lindhard}},\ }\href {http://publ.royalacademy.dk/books/414/2859} {\bibfield
  {journal} {\bibinfo  {journal} {Kgl. Danske Videnskab. Selskab Mat.-Fys.
  Medd.}\ }\textbf {\bibinfo {volume} {28}},\ \bibinfo {pages} {1} (\bibinfo
  {year} {1954})}\BibitemShut {NoStop}%
\bibitem [{\citenamefont {Mermin}(1970)}]{Mermin:1970}%
  \BibitemOpen
  \bibfield  {author} {\bibinfo {author} {\bibfnamefont {N.~D.}\ \bibnamefont
  {Mermin}},\ }\href {\doibase 10.1103/PhysRevB.1.2362} {\bibfield  {journal}
  {\bibinfo  {journal} {Phys. Rev. B}\ }\textbf {\bibinfo {volume} {1}},\
  \bibinfo {pages} {2362} (\bibinfo {year} {1970})}\BibitemShut {NoStop}%
\bibitem [{\citenamefont {Gallinet}\ \emph {et~al.}(2015)\citenamefont
  {Gallinet}, \citenamefont {Butet},\ and\ \citenamefont
  {Martin}}]{Gallinet:2015}%
  \BibitemOpen
  \bibfield  {author} {\bibinfo {author} {\bibfnamefont {B.}~\bibnamefont
  {Gallinet}}, \bibinfo {author} {\bibfnamefont {J.}~\bibnamefont {Butet}}, \
  and\ \bibinfo {author} {\bibfnamefont {O.~J.~F.}\ \bibnamefont {Martin}},\
  }\href {\doibase 10.1002/lpor.201500122} {\bibfield  {journal} {\bibinfo
  {journal} {Laser Photon. Rev.}\ }\textbf {\bibinfo {volume} {9}},\ \bibinfo
  {pages} {577} (\bibinfo {year} {2015})}\BibitemShut {NoStop}%
\bibitem [{\citenamefont {Joannopoulos}\ \emph {et~al.}(2008)\citenamefont
  {Joannopoulos}, \citenamefont {Johnson}, \citenamefont {Winn},\ and\
  \citenamefont {Meade}}]{Joannopoulos:2008}%
  \BibitemOpen
  \bibfield  {author} {\bibinfo {author} {\bibfnamefont {J.}~\bibnamefont
  {Joannopoulos}}, \bibinfo {author} {\bibfnamefont {S.}~\bibnamefont
  {Johnson}}, \bibinfo {author} {\bibfnamefont {J.}~\bibnamefont {Winn}}, \
  and\ \bibinfo {author} {\bibfnamefont {R.}~\bibnamefont {Meade}},\ }\href
  {http://ab-initio.mit.edu/book/photonic-crystals-book.pdf} {\emph {\bibinfo
  {title} {Photonic Crystals: Molding the Flow of Light (Second Edition)}}}\
  (\bibinfo  {publisher} {Princeton University Press},\ \bibinfo {year}
  {2008})\BibitemShut {NoStop}%
\bibitem [{\citenamefont {Dombi}\ \emph {et~al.}(2020)\citenamefont {Dombi},
  \citenamefont {P\'apa}, \citenamefont {Vogelsang}, \citenamefont {Yalunin},
  \citenamefont {Sivis}, \citenamefont {Herink}, \citenamefont {Sch\"afer},
  \citenamefont {Gro\ss{}}, \citenamefont {Ropers},\ and\ \citenamefont
  {Lienau}}]{Dombi:2020}%
  \BibitemOpen
  \bibfield  {author} {\bibinfo {author} {\bibfnamefont {P.}~\bibnamefont
  {Dombi}}, \bibinfo {author} {\bibfnamefont {Z.}~\bibnamefont {P\'apa}},
  \bibinfo {author} {\bibfnamefont {J.}~\bibnamefont {Vogelsang}}, \bibinfo
  {author} {\bibfnamefont {S.~V.}\ \bibnamefont {Yalunin}}, \bibinfo {author}
  {\bibfnamefont {M.}~\bibnamefont {Sivis}}, \bibinfo {author} {\bibfnamefont
  {G.}~\bibnamefont {Herink}}, \bibinfo {author} {\bibfnamefont
  {S.}~\bibnamefont {Sch\"afer}}, \bibinfo {author} {\bibfnamefont
  {P.}~\bibnamefont {Gro\ss{}}}, \bibinfo {author} {\bibfnamefont
  {C.}~\bibnamefont {Ropers}}, \ and\ \bibinfo {author} {\bibfnamefont
  {C.}~\bibnamefont {Lienau}},\ }\href {\doibase 10.1103/RevModPhys.92.025003}
  {\bibfield  {journal} {\bibinfo  {journal} {Rev. Mod. Phys.}\ }\textbf
  {\bibinfo {volume} {92}},\ \bibinfo {pages} {025003} (\bibinfo {year}
  {2020})}\BibitemShut {NoStop}%
\bibitem [{\citenamefont {Liebsch}(1987)}]{Liebsch:1987}%
  \BibitemOpen
  \bibfield  {author} {\bibinfo {author} {\bibfnamefont {A.}~\bibnamefont
  {Liebsch}},\ }\href {\doibase 10.1103/PhysRevB.36.7378} {\bibfield  {journal}
  {\bibinfo  {journal} {Phys. Rev. B}\ }\textbf {\bibinfo {volume} {36}},\
  \bibinfo {pages} {7378} (\bibinfo {year} {1987})}\BibitemShut {NoStop}%
\bibitem [{\citenamefont {Apell}(1981)}]{Apell:1981}%
  \BibitemOpen
  \bibfield  {author} {\bibinfo {author} {\bibfnamefont {P.}~\bibnamefont
  {Apell}},\ }\href {\doibase 10.1088/0031-8949/24/4/019} {\bibfield  {journal}
  {\bibinfo  {journal} {Phys. Scr.}\ }\textbf {\bibinfo {volume} {24}},\
  \bibinfo {pages} {795} (\bibinfo {year} {1981})}\BibitemShut {NoStop}%
\bibitem [{\citenamefont {Forstmann}\ and\ \citenamefont
  {Gerhardts}(1986)}]{Forstmann:1986}%
  \BibitemOpen
  \bibfield  {author} {\bibinfo {author} {\bibfnamefont {F.}~\bibnamefont
  {Forstmann}}\ and\ \bibinfo {author} {\bibfnamefont {R.~R.}\ \bibnamefont
  {Gerhardts}},\ }\href {\doibase 10.1007/BFb0048841} {\emph {\bibinfo {title}
  {Metal Optics Near the Plasma Frequency}}}\ (\bibinfo  {publisher}
  {Springer-Verlag Berlin Heidelberg},\ \bibinfo {year} {1986})\BibitemShut
  {NoStop}%
\bibitem [{\citenamefont {Langreth}(1989)}]{Langreth:1989}%
  \BibitemOpen
  \bibfield  {author} {\bibinfo {author} {\bibfnamefont {D.~C.}\ \bibnamefont
  {Langreth}},\ }\href {\doibase 10.1103/PhysRevB.39.10020} {\bibfield
  {journal} {\bibinfo  {journal} {Phys. Rev. B}\ }\textbf {\bibinfo {volume}
  {39}},\ \bibinfo {pages} {10020} (\bibinfo {year} {1989})}\BibitemShut
  {NoStop}%
\bibitem [{\citenamefont {Bagchi}\ \emph {et~al.}(1979)\citenamefont {Bagchi},
  \citenamefont {Barrera},\ and\ \citenamefont {Rajagopal}}]{Bagchi:1979}%
  \BibitemOpen
  \bibfield  {author} {\bibinfo {author} {\bibfnamefont {A.}~\bibnamefont
  {Bagchi}}, \bibinfo {author} {\bibfnamefont {R.~G.}\ \bibnamefont {Barrera}},
  \ and\ \bibinfo {author} {\bibfnamefont {A.~K.}\ \bibnamefont {Rajagopal}},\
  }\href {\doibase 10.1103/PhysRevB.20.4824} {\bibfield  {journal} {\bibinfo
  {journal} {Phys. Rev. B}\ }\textbf {\bibinfo {volume} {20}},\ \bibinfo
  {pages} {4824} (\bibinfo {year} {1979})}\BibitemShut {NoStop}%
\bibitem [{\citenamefont {Feibelman}(1981)}]{Feibelman:1981}%
  \BibitemOpen
  \bibfield  {author} {\bibinfo {author} {\bibfnamefont {P.~J.}\ \bibnamefont
  {Feibelman}},\ }\href {\doibase 10.1103/PhysRevB.23.2629} {\bibfield
  {journal} {\bibinfo  {journal} {Phys. Rev. B}\ }\textbf {\bibinfo {volume}
  {23}},\ \bibinfo {pages} {2629} (\bibinfo {year} {1981})}\BibitemShut
  {NoStop}%
\bibitem [{\citenamefont {Apell}(1982)}]{Apell:1982}%
  \BibitemOpen
  \bibfield  {author} {\bibinfo {author} {\bibfnamefont {P.}~\bibnamefont
  {Apell}},\ }\href {\doibase 10.1088/0031-8949/25/1a/009} {\bibfield
  {journal} {\bibinfo  {journal} {Phys. Scr.}\ }\textbf {\bibinfo {volume}
  {25}},\ \bibinfo {pages} {57} (\bibinfo {year} {1982})}\BibitemShut {NoStop}%
\bibitem [{\citenamefont {Apell}(1983)}]{Apell:1983a}%
  \BibitemOpen
  \bibfield  {author} {\bibinfo {author} {\bibfnamefont {P.}~\bibnamefont
  {Apell}},\ }\href {\doibase 10.1016/0038-1098(83)90763-9} {\bibfield
  {journal} {\bibinfo  {journal} {Solid State Commun.}\ }\textbf {\bibinfo
  {volume} {47}},\ \bibinfo {pages} {619} (\bibinfo {year} {1983})}\BibitemShut
  {NoStop}%
\bibitem [{\citenamefont {Ahlqvist}\ and\ \citenamefont
  {Apell}(1982)}]{Ahlqvist:1982}%
  \BibitemOpen
  \bibfield  {author} {\bibinfo {author} {\bibfnamefont {P.}~\bibnamefont
  {Ahlqvist}}\ and\ \bibinfo {author} {\bibfnamefont {P.}~\bibnamefont
  {Apell}},\ }\href {\doibase 10.1088/0031-8949/25/4/016} {\bibfield  {journal}
  {\bibinfo  {journal} {Phys. Scr.}\ }\textbf {\bibinfo {volume} {25}},\
  \bibinfo {pages} {587} (\bibinfo {year} {1982})}\BibitemShut {NoStop}%
\bibitem [{\citenamefont {Bagchi}\ \emph {et~al.}(1978)\citenamefont {Bagchi},
  \citenamefont {Kar},\ and\ \citenamefont {Barrera}}]{Bagchi:1978}%
  \BibitemOpen
  \bibfield  {author} {\bibinfo {author} {\bibfnamefont {A.}~\bibnamefont
  {Bagchi}}, \bibinfo {author} {\bibfnamefont {N.}~\bibnamefont {Kar}}, \ and\
  \bibinfo {author} {\bibfnamefont {R.~G.}\ \bibnamefont {Barrera}},\ }\href
  {\doibase 10.1103/PhysRevLett.40.803} {\bibfield  {journal} {\bibinfo
  {journal} {Phys. Rev. Lett.}\ }\textbf {\bibinfo {volume} {40}},\ \bibinfo
  {pages} {803} (\bibinfo {year} {1978})}\BibitemShut {NoStop}%
\bibitem [{\citenamefont {Liu}\ \emph {et~al.}(2017)\citenamefont {Liu},
  \citenamefont {Kang}, \citenamefont {Yuan}, \citenamefont {Park},
  \citenamefont {Kim}, \citenamefont {Cui}, \citenamefont {Hwang},\ and\
  \citenamefont {Brongersma}}]{Liu:2017}%
  \BibitemOpen
  \bibfield  {author} {\bibinfo {author} {\bibfnamefont {X.}~\bibnamefont
  {Liu}}, \bibinfo {author} {\bibfnamefont {H.}~\bibnamefont {Kang}}, \bibinfo
  {author} {\bibfnamefont {H.}~\bibnamefont {Yuan}}, \bibinfo {author}
  {\bibfnamefont {J.}~\bibnamefont {Park}}, \bibinfo {author} {\bibfnamefont
  {S.~J.}\ \bibnamefont {Kim}}, \bibinfo {author} {\bibfnamefont
  {Y.}~\bibnamefont {Cui}}, \bibinfo {author} {\bibfnamefont {H.~Y.}\
  \bibnamefont {Hwang}}, \ and\ \bibinfo {author} {\bibfnamefont {M.~L.}\
  \bibnamefont {Brongersma}},\ }\href {\doibase 10.1038/NNANO.2017.103}
  {\bibfield  {journal} {\bibinfo  {journal} {Nat. Nanotechnol.}\ }\textbf
  {\bibinfo {volume} {12}},\ \bibinfo {pages} {866} (\bibinfo {year}
  {2017})}\BibitemShut {NoStop}%
\bibitem [{\citenamefont {Skj\o{}lstrup}\ \emph {et~al.}(2018)\citenamefont
  {Skj\o{}lstrup}, \citenamefont {S\o{}ndergaard},\ and\ \citenamefont
  {Pedersen}}]{Skjolstrup:2018}%
  \BibitemOpen
  \bibfield  {author} {\bibinfo {author} {\bibfnamefont {E.~J.~H.}\
  \bibnamefont {Skj\o{}lstrup}}, \bibinfo {author} {\bibfnamefont
  {T.}~\bibnamefont {S\o{}ndergaard}}, \ and\ \bibinfo {author} {\bibfnamefont
  {T.~G.}\ \bibnamefont {Pedersen}},\ }\href {\doibase
  10.1103/PhysRevB.97.115429} {\bibfield  {journal} {\bibinfo  {journal} {Phys.
  Rev. B}\ }\textbf {\bibinfo {volume} {97}},\ \bibinfo {pages} {115429}
  (\bibinfo {year} {2018})}\BibitemShut {NoStop}%
\bibitem [{\citenamefont {Skj\o{}lstrup}\ \emph {et~al.}(2019)\citenamefont
  {Skj\o{}lstrup}, \citenamefont {S\o{}ndergaard},\ and\ \citenamefont
  {Pedersen}}]{Skjolstrup:2019}%
  \BibitemOpen
  \bibfield  {author} {\bibinfo {author} {\bibfnamefont {E.~J.~H.}\
  \bibnamefont {Skj\o{}lstrup}}, \bibinfo {author} {\bibfnamefont
  {T.}~\bibnamefont {S\o{}ndergaard}}, \ and\ \bibinfo {author} {\bibfnamefont
  {T.~G.}\ \bibnamefont {Pedersen}},\ }\href {\doibase
  10.1103/PhysRevB.99.155427} {\bibfield  {journal} {\bibinfo  {journal} {Phys.
  Rev. B}\ }\textbf {\bibinfo {volume} {99}},\ \bibinfo {pages} {155427}
  (\bibinfo {year} {2019})}\BibitemShut {NoStop}%
\bibitem [{\citenamefont {Taghizadeh}\ and\ \citenamefont
  {Pedersen}(2019)}]{Taghizadeh:2019}%
  \BibitemOpen
  \bibfield  {author} {\bibinfo {author} {\bibfnamefont {A.}~\bibnamefont
  {Taghizadeh}}\ and\ \bibinfo {author} {\bibfnamefont {T.~G.}\ \bibnamefont
  {Pedersen}},\ }\href {\doibase 10.1364/OE.27.036941} {\bibfield  {journal}
  {\bibinfo  {journal} {Opt. Express}\ }\textbf {\bibinfo {volume} {27}},\
  \bibinfo {pages} {36941} (\bibinfo {year} {2019})}\BibitemShut {NoStop}%
\bibitem [{\citenamefont {Rivacoba}(2019)}]{Rivacoba:2019}%
  \BibitemOpen
  \bibfield  {author} {\bibinfo {author} {\bibfnamefont {A.}~\bibnamefont
  {Rivacoba}},\ }\href {\doibase 10.1016/j.ultramic.2019.112835} {\bibfield
  {journal} {\bibinfo  {journal} {Ultramicroscopy}\ }\textbf {\bibinfo {volume}
  {207}},\ \bibinfo {pages} {112835} (\bibinfo {year} {2019})}\BibitemShut
  {NoStop}%
\bibitem [{\citenamefont {Rogowska}\ \emph {et~al.}(1994)\citenamefont
  {Rogowska}, \citenamefont {Wojciechowski},\ and\ \citenamefont
  {Maciejewski}}]{Rogowska:1994}%
  \BibitemOpen
  \bibfield  {author} {\bibinfo {author} {\bibfnamefont {J.~M.}\ \bibnamefont
  {Rogowska}}, \bibinfo {author} {\bibfnamefont {K.~F.}\ \bibnamefont
  {Wojciechowski}}, \ and\ \bibinfo {author} {\bibfnamefont {M.}~\bibnamefont
  {Maciejewski}},\ }\href {\doibase 10.12693/APhysPolA.85.593} {\bibfield
  {journal} {\bibinfo  {journal} {Acta Phys. Pol. A}\ }\textbf {\bibinfo
  {volume} {85}},\ \bibinfo {pages} {593} (\bibinfo {year} {1994})}\BibitemShut
  {NoStop}%
\bibitem [{Note1()}]{Note1}%
  \BibitemOpen
  \bibinfo {note} {We note, however, that this might not be the case for polar
  materials near its optical phonon frequencies.}\BibitemShut {Stop}%
\bibitem [{\citenamefont {Dethe}\ \emph {et~al.}(2019)\citenamefont {Dethe},
  \citenamefont {Gill}, \citenamefont {Green}, \citenamefont {Greensweight},
  \citenamefont {Gutierrez}, \citenamefont {He}, \citenamefont {Tajima},\ and\
  \citenamefont {Yang}}]{Dethe:2019}%
  \BibitemOpen
  \bibfield  {author} {\bibinfo {author} {\bibfnamefont {T.}~\bibnamefont
  {Dethe}}, \bibinfo {author} {\bibfnamefont {H.}~\bibnamefont {Gill}},
  \bibinfo {author} {\bibfnamefont {D.}~\bibnamefont {Green}}, \bibinfo
  {author} {\bibfnamefont {A.}~\bibnamefont {Greensweight}}, \bibinfo {author}
  {\bibfnamefont {L.}~\bibnamefont {Gutierrez}}, \bibinfo {author}
  {\bibfnamefont {M.}~\bibnamefont {He}}, \bibinfo {author} {\bibfnamefont
  {T.}~\bibnamefont {Tajima}}, \ and\ \bibinfo {author} {\bibfnamefont
  {K.}~\bibnamefont {Yang}},\ }\href {\doibase 10.1119/1.5092679} {\bibfield
  {journal} {\bibinfo  {journal} {Am. J. Phys.}\ }\textbf {\bibinfo {volume}
  {87}},\ \bibinfo {pages} {279} (\bibinfo {year} {2019})}\BibitemShut
  {NoStop}%
\bibitem [{\citenamefont {Persson}\ and\ \citenamefont
  {Apell}(1983)}]{Persson:1983}%
  \BibitemOpen
  \bibfield  {author} {\bibinfo {author} {\bibfnamefont {B.~N.~J.}\
  \bibnamefont {Persson}}\ and\ \bibinfo {author} {\bibfnamefont
  {P.}~\bibnamefont {Apell}},\ }\href {\doibase 10.1103/PhysRevB.27.6058}
  {\bibfield  {journal} {\bibinfo  {journal} {Phys. Rev. B}\ }\textbf {\bibinfo
  {volume} {27}},\ \bibinfo {pages} {6058} (\bibinfo {year}
  {1983})}\BibitemShut {NoStop}%
\bibitem [{\citenamefont {Persson}\ and\ \citenamefont
  {Zaremba}(1985)}]{Persson:1985}%
  \BibitemOpen
  \bibfield  {author} {\bibinfo {author} {\bibfnamefont {B.~N.~J.}\
  \bibnamefont {Persson}}\ and\ \bibinfo {author} {\bibfnamefont
  {E.}~\bibnamefont {Zaremba}},\ }\href {\doibase 10.1103/PhysRevB.31.1863}
  {\bibfield  {journal} {\bibinfo  {journal} {Phys. Rev. B}\ }\textbf {\bibinfo
  {volume} {31}},\ \bibinfo {pages} {1863} (\bibinfo {year}
  {1985})}\BibitemShut {NoStop}%
\bibitem [{\citenamefont {Lang}\ and\ \citenamefont {Kohn}(1973)}]{Lang:1973}%
  \BibitemOpen
  \bibfield  {author} {\bibinfo {author} {\bibfnamefont {N.~D.}\ \bibnamefont
  {Lang}}\ and\ \bibinfo {author} {\bibfnamefont {W.}~\bibnamefont {Kohn}},\
  }\href {\doibase 10.1103/PhysRevB.7.3541} {\bibfield  {journal} {\bibinfo
  {journal} {Phys. Rev. B}\ }\textbf {\bibinfo {volume} {7}},\ \bibinfo {pages}
  {3541} (\bibinfo {year} {1973})}\BibitemShut {NoStop}%
\bibitem [{\citenamefont {Gon\c{c}alves}\ \emph
  {et~al.}(2020{\natexlab{b}})\citenamefont {Gon\c{c}alves}, \citenamefont
  {Stenger}, \citenamefont {Cox}, \citenamefont {Mortensen},\ and\
  \citenamefont {Xiao}}]{Goncalves:2020b}%
  \BibitemOpen
  \bibfield  {author} {\bibinfo {author} {\bibfnamefont {P.~A.~D.}\
  \bibnamefont {Gon\c{c}alves}}, \bibinfo {author} {\bibfnamefont
  {N.}~\bibnamefont {Stenger}}, \bibinfo {author} {\bibfnamefont {J.~D.}\
  \bibnamefont {Cox}}, \bibinfo {author} {\bibfnamefont {N.~A.}\ \bibnamefont
  {Mortensen}}, \ and\ \bibinfo {author} {\bibfnamefont {S.}~\bibnamefont
  {Xiao}},\ }\href {\doibase 10.1002/adom.201901473} {\bibfield  {journal}
  {\bibinfo  {journal} {Adv. Opt. Mater.}\ }\textbf {\bibinfo {volume} {8}},\
  \bibinfo {pages} {1901473} (\bibinfo {year}
  {2020}{\natexlab{b}})}\BibitemShut {NoStop}%
\bibitem [{\citenamefont {Reserbat-Plantey}\ \emph {et~al.}(2021)\citenamefont
  {Reserbat-Plantey}, \citenamefont {Epstein}, \citenamefont {Torre},
  \citenamefont {Costa}, \citenamefont {Gon\c{c}alves}, \citenamefont
  {Mortensen}, \citenamefont {Polini}, \citenamefont {Song}, \citenamefont
  {Peres},\ and\ \citenamefont {Koppens}}]{Reserbat-Plantey:2021}%
  \BibitemOpen
  \bibfield  {author} {\bibinfo {author} {\bibfnamefont {A.}~\bibnamefont
  {Reserbat-Plantey}}, \bibinfo {author} {\bibfnamefont {I.}~\bibnamefont
  {Epstein}}, \bibinfo {author} {\bibfnamefont {I.}~\bibnamefont {Torre}},
  \bibinfo {author} {\bibfnamefont {A.~T.}\ \bibnamefont {Costa}}, \bibinfo
  {author} {\bibfnamefont {P.~A.~D.}\ \bibnamefont {Gon\c{c}alves}}, \bibinfo
  {author} {\bibfnamefont {N.~A.}\ \bibnamefont {Mortensen}}, \bibinfo {author}
  {\bibfnamefont {M.}~\bibnamefont {Polini}}, \bibinfo {author} {\bibfnamefont
  {J.~C.~W.}\ \bibnamefont {Song}}, \bibinfo {author} {\bibfnamefont
  {N.~M.~R.}\ \bibnamefont {Peres}}, \ and\ \bibinfo {author} {\bibfnamefont
  {F.~H.~L.}\ \bibnamefont {Koppens}},\ }\href {\doibase
  10.1021/acsphotonics.0c01224} {\bibfield  {journal} {\bibinfo  {journal} {ACS
  Photonics}\ }\textbf {\bibinfo {volume} {8}},\ \bibinfo {pages} {85}
  (\bibinfo {year} {2021})}\BibitemShut {NoStop}%
\bibitem [{\citenamefont {Christensen}\ \emph {et~al.}(2014)\citenamefont
  {Christensen}, \citenamefont {Yan}, \citenamefont {Raza}, \citenamefont
  {Jauho}, \citenamefont {Mortensen},\ and\ \citenamefont
  {Wubs}}]{Christensen:2014}%
  \BibitemOpen
  \bibfield  {author} {\bibinfo {author} {\bibfnamefont {T.}~\bibnamefont
  {Christensen}}, \bibinfo {author} {\bibfnamefont {W.}~\bibnamefont {Yan}},
  \bibinfo {author} {\bibfnamefont {S.}~\bibnamefont {Raza}}, \bibinfo {author}
  {\bibfnamefont {A.-P.}\ \bibnamefont {Jauho}}, \bibinfo {author}
  {\bibfnamefont {N.~A.}\ \bibnamefont {Mortensen}}, \ and\ \bibinfo {author}
  {\bibfnamefont {M.}~\bibnamefont {Wubs}},\ }\href {\doibase
  10.1021/nn406153k} {\bibfield  {journal} {\bibinfo  {journal} {ACS Nano}\
  }\textbf {\bibinfo {volume} {8}},\ \bibinfo {pages} {1745} (\bibinfo {year}
  {2014})}\BibitemShut {NoStop}%
\bibitem [{\citenamefont {Wang}\ and\ \citenamefont {Shen}(2006)}]{Wang:2006}%
  \BibitemOpen
  \bibfield  {author} {\bibinfo {author} {\bibfnamefont {F.}~\bibnamefont
  {Wang}}\ and\ \bibinfo {author} {\bibfnamefont {Y.~R.}\ \bibnamefont
  {Shen}},\ }\href {\doibase 10.1103/PhysRevLett.97.206806} {\bibfield
  {journal} {\bibinfo  {journal} {Phys. Rev. Lett.}\ }\textbf {\bibinfo
  {volume} {97}},\ \bibinfo {pages} {206806} (\bibinfo {year}
  {2006})}\BibitemShut {NoStop}%
\bibitem [{\citenamefont {Bennett}(1970)}]{Bennett:1970}%
  \BibitemOpen
  \bibfield  {author} {\bibinfo {author} {\bibfnamefont {A.~J.}\ \bibnamefont
  {Bennett}},\ }\href {\doibase 10.1103/PhysRevB.1.203} {\bibfield  {journal}
  {\bibinfo  {journal} {Phys. Rev. B}\ }\textbf {\bibinfo {volume} {1}},\
  \bibinfo {pages} {203} (\bibinfo {year} {1970})}\BibitemShut {NoStop}%
\bibitem [{\citenamefont {Tsuei}\ \emph {et~al.}(1990)\citenamefont {Tsuei},
  \citenamefont {Plummer}, \citenamefont {Liebsch}, \citenamefont {Kempa},\
  and\ \citenamefont {Bakshi}}]{Tsuei:1990}%
  \BibitemOpen
  \bibfield  {author} {\bibinfo {author} {\bibfnamefont {K.-D.}\ \bibnamefont
  {Tsuei}}, \bibinfo {author} {\bibfnamefont {E.~W.}\ \bibnamefont {Plummer}},
  \bibinfo {author} {\bibfnamefont {A.}~\bibnamefont {Liebsch}}, \bibinfo
  {author} {\bibfnamefont {K.}~\bibnamefont {Kempa}}, \ and\ \bibinfo {author}
  {\bibfnamefont {P.}~\bibnamefont {Bakshi}},\ }\href {\doibase
  10.1103/PhysRevLett.64.44} {\bibfield  {journal} {\bibinfo  {journal} {Phys.
  Rev. Lett.}\ }\textbf {\bibinfo {volume} {64}},\ \bibinfo {pages} {44}
  (\bibinfo {year} {1990})}\BibitemShut {NoStop}%
\bibitem [{\citenamefont {Campos}\ \emph {et~al.}(2019)\citenamefont {Campos},
  \citenamefont {Troc}, \citenamefont {Cottancin}, \citenamefont {Pellarin},
  \citenamefont {Weissker}, \citenamefont {Lerm{\'e}}, \citenamefont {Kociak},\
  and\ \citenamefont {Hillenkamp}}]{Campos:2019}%
  \BibitemOpen
  \bibfield  {author} {\bibinfo {author} {\bibfnamefont {A.}~\bibnamefont
  {Campos}}, \bibinfo {author} {\bibfnamefont {N.}~\bibnamefont {Troc}},
  \bibinfo {author} {\bibfnamefont {E.}~\bibnamefont {Cottancin}}, \bibinfo
  {author} {\bibfnamefont {M.}~\bibnamefont {Pellarin}}, \bibinfo {author}
  {\bibfnamefont {H.-C.}\ \bibnamefont {Weissker}}, \bibinfo {author}
  {\bibfnamefont {J.}~\bibnamefont {Lerm{\'e}}}, \bibinfo {author}
  {\bibfnamefont {M.}~\bibnamefont {Kociak}}, \ and\ \bibinfo {author}
  {\bibfnamefont {M.}~\bibnamefont {Hillenkamp}},\ }\href {\doibase
  10.1038/s41567-018-0345-z} {\bibfield  {journal} {\bibinfo  {journal} {Nat.
  Phys.}\ }\textbf {\bibinfo {volume} {15}},\ \bibinfo {pages} {275} (\bibinfo
  {year} {2019})}\BibitemShut {NoStop}%
\bibitem [{\citenamefont {Liebsch}(1993)}]{Liebsch:1993}%
  \BibitemOpen
  \bibfield  {author} {\bibinfo {author} {\bibfnamefont {A.}~\bibnamefont
  {Liebsch}},\ }\href {\doibase 10.1103/PhysRevB.48.11317} {\bibfield
  {journal} {\bibinfo  {journal} {Phys. Rev. B}\ }\textbf {\bibinfo {volume}
  {48}},\ \bibinfo {pages} {11317} (\bibinfo {year} {1993})}\BibitemShut
  {NoStop}%
\bibitem [{\citenamefont {Scholl}\ \emph {et~al.}(2012)\citenamefont {Scholl},
  \citenamefont {Koh},\ and\ \citenamefont {Dionne}}]{Scholl:2012}%
  \BibitemOpen
  \bibfield  {author} {\bibinfo {author} {\bibfnamefont {J.~A.}\ \bibnamefont
  {Scholl}}, \bibinfo {author} {\bibfnamefont {A.~L.}\ \bibnamefont {Koh}}, \
  and\ \bibinfo {author} {\bibfnamefont {J.~A.}\ \bibnamefont {Dionne}},\
  }\href {\doibase 10.1038/nature10904} {\bibfield  {journal} {\bibinfo
  {journal} {Nature}\ }\textbf {\bibinfo {volume} {483}},\ \bibinfo {pages}
  {421} (\bibinfo {year} {2012})}\BibitemShut {NoStop}%
\bibitem [{\citenamefont {Halevi}(1995)}]{Halevi:1995}%
  \BibitemOpen
  \bibfield  {author} {\bibinfo {author} {\bibfnamefont {P.}~\bibnamefont
  {Halevi}},\ }\href {\doibase 10.1103/PhysRevB.51.7497} {\bibfield  {journal}
  {\bibinfo  {journal} {Phys. Rev. B}\ }\textbf {\bibinfo {volume} {51}},\
  \bibinfo {pages} {7497} (\bibinfo {year} {1995})}\BibitemShut {NoStop}%
\bibitem [{\citenamefont {Yan}(2015)}]{Yan:2015b}%
  \BibitemOpen
  \bibfield  {author} {\bibinfo {author} {\bibfnamefont {W.}~\bibnamefont
  {Yan}},\ }\href {\doibase 10.1103/PhysRevB.91.115416} {\bibfield  {journal}
  {\bibinfo  {journal} {Phys. Rev. B}\ }\textbf {\bibinfo {volume} {91}},\
  \bibinfo {pages} {115416} (\bibinfo {year} {2015})}\BibitemShut {NoStop}%
\bibitem [{\citenamefont {Cirac{\`i}}\ and\ \citenamefont
  {Della~Sala}(2016)}]{Ciraci:2016}%
  \BibitemOpen
  \bibfield  {author} {\bibinfo {author} {\bibfnamefont {C.}~\bibnamefont
  {Cirac{\`i}}}\ and\ \bibinfo {author} {\bibfnamefont {F.}~\bibnamefont
  {Della~Sala}},\ }\href {\doibase 10.1103/PhysRevB.93.205405} {\bibfield
  {journal} {\bibinfo  {journal} {Phys. Rev. B}\ }\textbf {\bibinfo {volume}
  {93}},\ \bibinfo {pages} {205405} (\bibinfo {year} {2016})}\BibitemShut
  {NoStop}%
\bibitem [{\citenamefont {Raza}\ \emph {et~al.}(2013)\citenamefont {Raza},
  \citenamefont {Stenger}, \citenamefont {Kadkhodazadeh}, \citenamefont
  {Fischer}, \citenamefont {Kostesha}, \citenamefont {Jauho}, \citenamefont
  {Burrows}, \citenamefont {Wubs},\ and\ \citenamefont
  {Mortensen}}]{Raza:2013}%
  \BibitemOpen
  \bibfield  {author} {\bibinfo {author} {\bibfnamefont {S.}~\bibnamefont
  {Raza}}, \bibinfo {author} {\bibfnamefont {N.}~\bibnamefont {Stenger}},
  \bibinfo {author} {\bibfnamefont {S.}~\bibnamefont {Kadkhodazadeh}}, \bibinfo
  {author} {\bibfnamefont {S.~V.}\ \bibnamefont {Fischer}}, \bibinfo {author}
  {\bibfnamefont {N.}~\bibnamefont {Kostesha}}, \bibinfo {author}
  {\bibfnamefont {A.-P.}\ \bibnamefont {Jauho}}, \bibinfo {author}
  {\bibfnamefont {A.}~\bibnamefont {Burrows}}, \bibinfo {author} {\bibfnamefont
  {M.}~\bibnamefont {Wubs}}, \ and\ \bibinfo {author} {\bibfnamefont {N.~A.}\
  \bibnamefont {Mortensen}},\ }\href {\doibase 10.1515/nanoph-2012-0032}
  {\bibfield  {journal} {\bibinfo  {journal} {Nanophotonics}\ }\textbf
  {\bibinfo {volume} {2}},\ \bibinfo {pages} {131} (\bibinfo {year}
  {2013})}\BibitemShut {NoStop}%
\bibitem [{\citenamefont {Raza}\ \emph
  {et~al.}(2015{\natexlab{b}})\citenamefont {Raza}, \citenamefont
  {Kadkhodazadeh}, \citenamefont {Christensen}, \citenamefont {{Di Vece}},
  \citenamefont {Wubs}, \citenamefont {Mortensen},\ and\ \citenamefont
  {Stenger}}]{Raza:2015}%
  \BibitemOpen
  \bibfield  {author} {\bibinfo {author} {\bibfnamefont {S.}~\bibnamefont
  {Raza}}, \bibinfo {author} {\bibfnamefont {S.}~\bibnamefont {Kadkhodazadeh}},
  \bibinfo {author} {\bibfnamefont {T.}~\bibnamefont {Christensen}}, \bibinfo
  {author} {\bibfnamefont {M.}~\bibnamefont {{Di Vece}}}, \bibinfo {author}
  {\bibfnamefont {M.}~\bibnamefont {Wubs}}, \bibinfo {author} {\bibfnamefont
  {N.~A.}\ \bibnamefont {Mortensen}}, \ and\ \bibinfo {author} {\bibfnamefont
  {N.}~\bibnamefont {Stenger}},\ }\href {\doibase 10.1038/ncomms9788}
  {\bibfield  {journal} {\bibinfo  {journal} {Nat. Commun.}\ }\textbf {\bibinfo
  {volume} {6}},\ \bibinfo {pages} {8788} (\bibinfo {year}
  {2015}{\natexlab{b}})}\BibitemShut {NoStop}%
\bibitem [{\citenamefont {Lundeberg}\ \emph {et~al.}(2017)\citenamefont
  {Lundeberg}, \citenamefont {Gao}, \citenamefont {Asgari}, \citenamefont
  {Tan}, \citenamefont {Van~Duppen}, \citenamefont {Autore}, \citenamefont
  {Alonso-Gonz{\'a}lez}, \citenamefont {Woessner}, \citenamefont {Watanabe},
  \citenamefont {Taniguchi}, \citenamefont {Hillenbrand}, \citenamefont {Hone},
  \citenamefont {Polini},\ and\ \citenamefont {Koppens}}]{Lundeberg:2017}%
  \BibitemOpen
  \bibfield  {author} {\bibinfo {author} {\bibfnamefont {M.~B.}\ \bibnamefont
  {Lundeberg}}, \bibinfo {author} {\bibfnamefont {Y.}~\bibnamefont {Gao}},
  \bibinfo {author} {\bibfnamefont {R.}~\bibnamefont {Asgari}}, \bibinfo
  {author} {\bibfnamefont {C.}~\bibnamefont {Tan}}, \bibinfo {author}
  {\bibfnamefont {B.}~\bibnamefont {Van~Duppen}}, \bibinfo {author}
  {\bibfnamefont {M.}~\bibnamefont {Autore}}, \bibinfo {author} {\bibfnamefont
  {P.}~\bibnamefont {Alonso-Gonz{\'a}lez}}, \bibinfo {author} {\bibfnamefont
  {A.}~\bibnamefont {Woessner}}, \bibinfo {author} {\bibfnamefont
  {K.}~\bibnamefont {Watanabe}}, \bibinfo {author} {\bibfnamefont
  {T.}~\bibnamefont {Taniguchi}}, \bibinfo {author} {\bibfnamefont
  {R.}~\bibnamefont {Hillenbrand}}, \bibinfo {author} {\bibfnamefont
  {J.}~\bibnamefont {Hone}}, \bibinfo {author} {\bibfnamefont {M.}~\bibnamefont
  {Polini}}, \ and\ \bibinfo {author} {\bibfnamefont {F.~H.~L.}\ \bibnamefont
  {Koppens}},\ }\href {\doibase 10.1126/science.aan2735} {\bibfield  {journal}
  {\bibinfo  {journal} {Science}\ }\textbf {\bibinfo {volume} {357}},\ \bibinfo
  {pages} {187} (\bibinfo {year} {2017})}\BibitemShut {NoStop}%
\bibitem [{\citenamefont {Iranzo}\ \emph {et~al.}(2018)\citenamefont {Iranzo},
  \citenamefont {Nanot}, \citenamefont {Dias}, \citenamefont {Epstein},
  \citenamefont {Peng}, \citenamefont {Efetov}, \citenamefont {Lundeberg},
  \citenamefont {Parret}, \citenamefont {Osmond}, \citenamefont {Hong},
  \citenamefont {Kong}, \citenamefont {Englund}, \citenamefont {Peres},\ and\
  \citenamefont {Koppens}}]{Iranzo:2018}%
  \BibitemOpen
  \bibfield  {author} {\bibinfo {author} {\bibfnamefont {D.~A.}\ \bibnamefont
  {Iranzo}}, \bibinfo {author} {\bibfnamefont {S.}~\bibnamefont {Nanot}},
  \bibinfo {author} {\bibfnamefont {E.~J.~C.}\ \bibnamefont {Dias}}, \bibinfo
  {author} {\bibfnamefont {I.}~\bibnamefont {Epstein}}, \bibinfo {author}
  {\bibfnamefont {C.}~\bibnamefont {Peng}}, \bibinfo {author} {\bibfnamefont
  {D.~K.}\ \bibnamefont {Efetov}}, \bibinfo {author} {\bibfnamefont {M.~B.}\
  \bibnamefont {Lundeberg}}, \bibinfo {author} {\bibfnamefont {R.}~\bibnamefont
  {Parret}}, \bibinfo {author} {\bibfnamefont {J.}~\bibnamefont {Osmond}},
  \bibinfo {author} {\bibfnamefont {J.-Y.}\ \bibnamefont {Hong}}, \bibinfo
  {author} {\bibfnamefont {J.}~\bibnamefont {Kong}}, \bibinfo {author}
  {\bibfnamefont {D.~R.}\ \bibnamefont {Englund}}, \bibinfo {author}
  {\bibfnamefont {N.~M.~R.}\ \bibnamefont {Peres}}, \ and\ \bibinfo {author}
  {\bibfnamefont {F.~H.~L.}\ \bibnamefont {Koppens}},\ }\href {\doibase
  10.1126/science.aar8438} {\bibfield  {journal} {\bibinfo  {journal}
  {Science}\ }\textbf {\bibinfo {volume} {360}},\ \bibinfo {pages} {291}
  (\bibinfo {year} {2018})}\BibitemShut {NoStop}%
\bibitem [{\citenamefont {Dias}\ \emph {et~al.}(2018)\citenamefont {Dias},
  \citenamefont {Iranzo}, \citenamefont {Gon\ifmmode~\mbox{\c{c}}\else
  \c{c}\fi{}alves}, \citenamefont {Hajati}, \citenamefont {Bludov},
  \citenamefont {Jauho}, \citenamefont {Mortensen}, \citenamefont {Koppens},\
  and\ \citenamefont {Peres}}]{Dias:2018}%
  \BibitemOpen
  \bibfield  {author} {\bibinfo {author} {\bibfnamefont {E.~J.~C.}\
  \bibnamefont {Dias}}, \bibinfo {author} {\bibfnamefont {D.~A.}\ \bibnamefont
  {Iranzo}}, \bibinfo {author} {\bibfnamefont {P.~A.~D.}\ \bibnamefont
  {Gon\ifmmode~\mbox{\c{c}}\else \c{c}\fi{}alves}}, \bibinfo {author}
  {\bibfnamefont {Y.}~\bibnamefont {Hajati}}, \bibinfo {author} {\bibfnamefont
  {Y.~V.}\ \bibnamefont {Bludov}}, \bibinfo {author} {\bibfnamefont {A.-P.}\
  \bibnamefont {Jauho}}, \bibinfo {author} {\bibfnamefont {N.~A.}\ \bibnamefont
  {Mortensen}}, \bibinfo {author} {\bibfnamefont {F.~H.~L.}\ \bibnamefont
  {Koppens}}, \ and\ \bibinfo {author} {\bibfnamefont {N.~M.~R.}\ \bibnamefont
  {Peres}},\ }\href {\doibase 10.1103/PhysRevB.97.245405} {\bibfield  {journal}
  {\bibinfo  {journal} {Phys. Rev. B}\ }\textbf {\bibinfo {volume} {97}},\
  \bibinfo {pages} {245405} (\bibinfo {year} {2018})}\BibitemShut {NoStop}%
\bibitem [{\citenamefont {Toscano}\ \emph {et~al.}(2012)\citenamefont
  {Toscano}, \citenamefont {Raza}, \citenamefont {Xiao}, \citenamefont {Wubs},
  \citenamefont {Jauho}, \citenamefont {Bozhevolnyi},\ and\ \citenamefont
  {Mortensen}}]{Toscano:2012b}%
  \BibitemOpen
  \bibfield  {author} {\bibinfo {author} {\bibfnamefont {G.}~\bibnamefont
  {Toscano}}, \bibinfo {author} {\bibfnamefont {S.}~\bibnamefont {Raza}},
  \bibinfo {author} {\bibfnamefont {S.}~\bibnamefont {Xiao}}, \bibinfo {author}
  {\bibfnamefont {M.}~\bibnamefont {Wubs}}, \bibinfo {author} {\bibfnamefont
  {A.-P.}\ \bibnamefont {Jauho}}, \bibinfo {author} {\bibfnamefont {S.~I.}\
  \bibnamefont {Bozhevolnyi}}, \ and\ \bibinfo {author} {\bibfnamefont {N.~A.}\
  \bibnamefont {Mortensen}},\ }\href {\doibase 10.1364/OL.37.002538} {\bibfield
   {journal} {\bibinfo  {journal} {Opt. Lett.}\ }\textbf {\bibinfo {volume}
  {37}},\ \bibinfo {pages} {2538} (\bibinfo {year} {2012})}\BibitemShut
  {NoStop}%
\bibitem [{\citenamefont {Larkin}\ and\ \citenamefont
  {Stockman}(2005)}]{Larkin:2005}%
  \BibitemOpen
  \bibfield  {author} {\bibinfo {author} {\bibfnamefont {I.~A.}\ \bibnamefont
  {Larkin}}\ and\ \bibinfo {author} {\bibfnamefont {M.~I.}\ \bibnamefont
  {Stockman}},\ }\href {\doibase 10.1021/nl047957a} {\bibfield  {journal}
  {\bibinfo  {journal} {Nano Lett.}\ }\textbf {\bibinfo {volume} {5}},\
  \bibinfo {pages} {339} (\bibinfo {year} {2005})}\BibitemShut {NoStop}%
\end{thebibliography}
\end{document}